\documentclass[graybox]{svmult}

\usepackage{type1cm}        
\usepackage{makeidx}         
\usepackage{graphicx}        
\usepackage{multicol}        
\usepackage[bottom]{footmisc}

\usepackage{newtxtext}       %
\usepackage{newtxmath}       

\usepackage[hidelinks]{hyperref}        
\usepackage{url}                        

\makeindex             

\begin{document}

\title*{Agent-Based Modeling and its Tradeoffs:  An Introduction \& Examples}
\titlerunning{Intro to ABM with Examples}
\author{G. Wade McDonald and Nathaniel D. Osgood}
\institute{G. Wade McDonald \at University of Saskatchewan, Saskatoon, SK, Canada, \email{gwm762@mail.usask.ca}
\and Nathaniel D. Osgood \at University of Saskatchewan, Saskatoon, SK, Canada \email{osgood@cs.usask.ca}}
%
%
\maketitle

\abstract*{Agent-based modeling is a computational dynamic modeling technique that may be less familiar to some readers. Agent-based modeling seeks to understand the behaviour of complex systems by situating agents in an environment and studying the emergent outcomes of agent-agent and agent-environment interactions. In comparison with compartmental models, agent-based models offer simpler, more scalable and flexible representation of heterogeneity, the ability to capture dynamic and static network and spatial context, and the ability to consider history of individuals within the model. In contrast, compartmental models offer faster development time with less programming required, lower computational requirements that do not scale with population, and the option for concise mathematical formulation with ordinary, delay or stochastic differential equations supporting derivation of properties of the system behaviour.
In this chapter, basic characteristics of agent-based models are introduced, advantages and disadvantages of agent-based models, as compared with compartmental models, are discussed, and two example agent-based infectious disease models are reviewed.
}

\abstract{Agent-based modeling is a computational dynamic modeling technique that may be less familiar to some readers. Agent-based modeling seeks to understand the behaviour of complex systems by situating agents in an environment and studying the emergent outcomes of agent-agent and agent-environment interactions. In comparison with compartmental models, agent-based models offer simpler, more scalable and flexible representation of heterogeneity, the ability to capture dynamic and static network and spatial context, and the ability to consider history of individuals within the model. In contrast, compartmental models offer faster development time with less programming required, lower computational requirements that do not scale with population, and the option for concise mathematical formulation with ordinary, delay or stochastic differential equations supporting derivation of properties of the system behaviour.\newline\indent
In this chapter, basic characteristics of agent-based models are introduced, advantages and disadvantages of agent-based models, as compared with compartmental models, are discussed, and two example agent-based infectious disease models are reviewed.
}

\section{Introduction}
\label{sec:intro}
Agent-Based Modeling (ABM) is a computational dynamic modeling technique which seeks to understand behaviour of complex systems through the lens of agent-agent and agent-environment interactions \cite{railsback2019agent}. Agent-based models (ABMs) can be said to be ``upwards-facing'' or ``bottom-up'' in the sense that we specify behaviour of situated agents and these agents interact, which dictates higher-level system behaviour. Patterns and often surprising results emerge over time, space, and networks, possibly at multiple levels of the system.

\section{Characteristics of Agent-Based Models}
\label{sec:characteristics}
The origins of Agent-Based Modeling can be traced back to von Neumann and Ulam's work in the 1940s on replicating and cellular automata \cite{neumann1966theory, ulam1972some}. Since that time, the tradition has been enriched by contributions from Computational Physics, Computer Science, Mathematics, and from microsimulation modeling in Economics.  ABMs, like other types of dynamic models, vary from the extremely stylized and simple thinking tools for theory building \cite{hammond2009peer} to descriptively rich and empirically grounded models that seek to support theory explication and understanding the logical consequences of theory over time \cite{hammond2009peer}. 

ABMs consist of one or more populations of agents, where each such agent is equipped with parameters (representing pre-specified assumptions), state (characterizing an underlying situation evolving over time), and actions that change that state according to some rules or rates of change. These models are specified over some time horizon according to  either a continuous or discrete time abstraction. \textit{Continuous time modeling} abstractions support discrete events occurring at real-valued times at whatever tempo, pace and temporal granularity is required for particular circumstances within the model. For example, such an event might be associated with each occurrence of infection, recovery, vaccination, contact, death or birth. By contrast, the \textit{discrete time abstraction} involves updates to model agents and environments in lockstep in monolithic (atomic) ``ticks'' or ``timesteps''; in the event that there are multiple processes that need to be considered within the timespan represented by a given such timestep, their effects need to be brokered in the associated update to model step. For example, each timestep might represent a month as a whole, and on reaching that timestep, all of the distinct processes occurring during that month (deaths, infections, births, vaccinations) would need to be considered.  

Beyond having properties of it own, each agent is situated in an environment, which can include static or dynamic networks, spatial context that may be geographic or stylized, and potentially several levels of context. Sometimes such environments are highly evidenced and empirically grounded. Beyond these, there are typically some outputs of model state or changes therein reported by the model; sometimes those are governing factors in the model, are often instead epiphenomenal -- that is, reporting on but not influencing model state evolution. Finally, interventions represent mechanisms that alter elements of a model to permit investigation of counterfactual scenarios.

\subsection{Parameters}
\label{sec:parameters}
\begin{figure}
\sidecaption
\includegraphics[width=\textwidth]{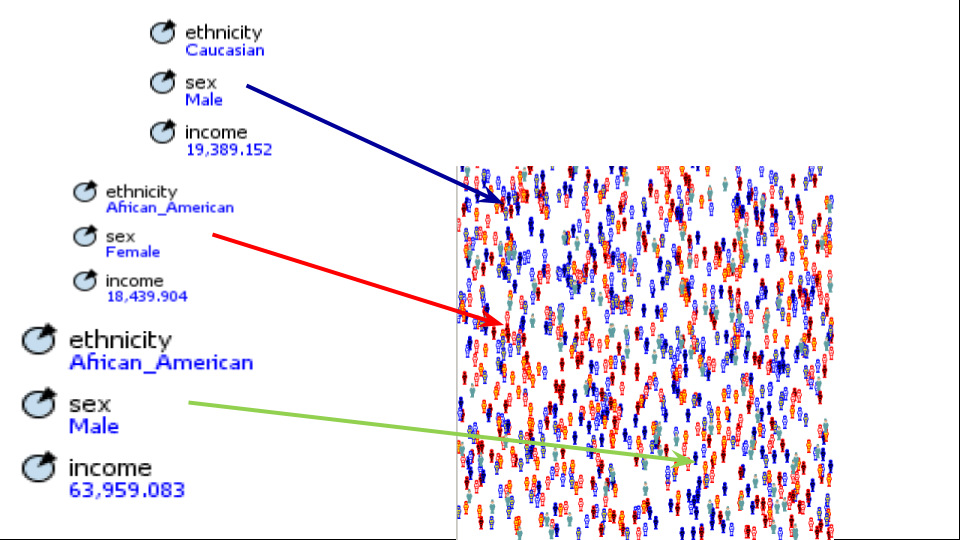}
\caption{An example of agent parameters: Parameter values are constant or otherwise pre-specified, but they do vary from agent-to-agent across a population.}
\label{fig:parameters}
\end{figure}

Figure \ref{fig:parameters} shows an example of a population of agents within an agent-based model, with each agent representing a person and having parameters for sex, income, and ethnicity. Values of these parameters are fixed over time but vary from agent-to-agent within the population. Agent-based models commonly include parameters that are continuous (e.g., income), discrete (e.g., sex) in character; they can also be relational -- such as a parameter referring to an agent's mother or school. In this context, \textit{continuous} refers to a parameter that may hold any value on an interval on the Real Number line (as approximated by floating-point arithmetic), while \textit{discrete} refers to a parameter that may take on a limited number of distinct values, whether numeric, ordinal or nominal. This capacity of agent-based models to capture continuous and relational heterogeneity stands in contrast to the fact that the only general approach to representing heterogeneity in aggregate models -- via model stratification -- is limited to representing discrete heterogeneity. A further advantage of individual-level representation -- the capacity to scale far more effectively than aggregate models as the number of types of heterogeneity rises, and to nimbly evolve the types of heterogeneity represented -- is also shared by aspects of state representation, and will be discussed below.

\subsection{State, Actions, and Rules}
\label{sec:state}
\begin{figure}[t]
\sidecaption[t]
\includegraphics[width=\textwidth]{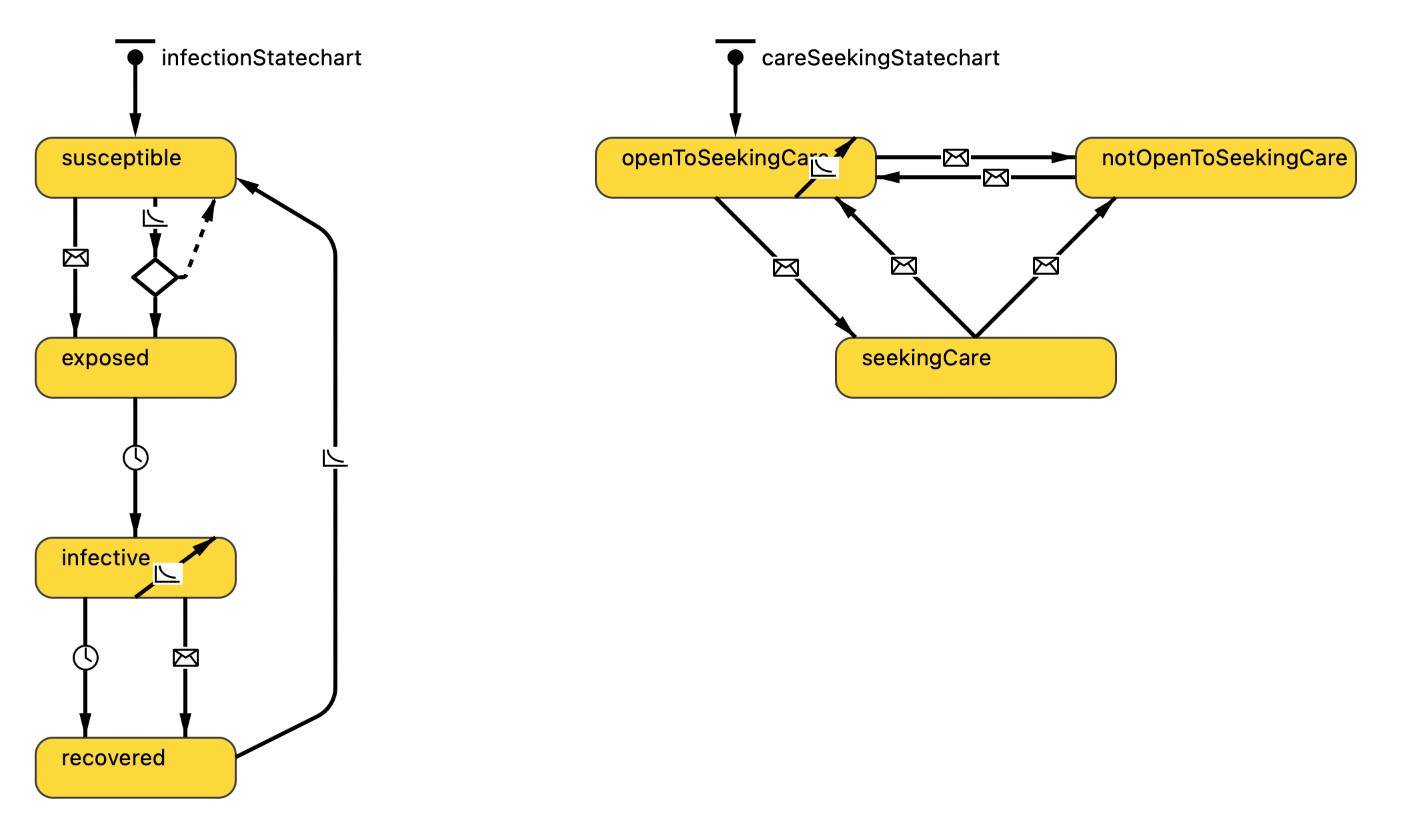}
\caption{Examples of statecharts in a single agent: Rounded rectangles represent states while arrows represent transitions, which encompass the actions and rules governing state change over time.}
\label{fig:statechart1}
\end{figure}

A number of ABM software packages --- including the AnyLogic software \cite{anylogic-software} used for the example models here --- describe agent state, and the actions and rules by which it evolves, using statecharts. Figure \ref{fig:statechart1} shows an example statechart for an agent representing a person in an infectious disease model. A person is associated with a set of possible states related to infection status indicating that at any one time they are either susceptible, exposed, infective, or recovered. Over time, they evolve between those states. The statechart at once depicts the possible states as rounded rectangles as well as the actions that can change state as arrows -- for example, transitioning from latent infection to infectiousness. The transition internal to the infective state is associated with this agent's exposure of other agents to pathogen. The iconography on the arrows hints to the fact that there are rules of various types governing these actions. Within one statechart, states are mutually exclusive and collectively exhaustive. Unless explicitly coupled, multiple statecharts within one agent evolve independently.

When compared to aggregate dynamic modeling methods, the representation of state in agent-based models confers some advantages. We can readily represent a given agent's state and its evolution over time with respect to more than one type of concern -- for example, we can represent a person as in a particular state with respect to infection and a particular state with respect to care-seeking. Such a representation avoids the `curse of dimensionality' that confronts stratified aggregate models as a result of the combinatorial explosion of possible states of different conditions. The count of such combinations rises geometrically with the number of dimensions being represented \cite{osgood2009Comorbidities, osgood2004Heterogeneity}. For the example in Figure \ref{fig:statechart1}, there are two statecharts with a total of seven states; representation of the same level of heterogeneity using aggregate compartmental modeling techniques would require $4 \times 3 = 12$ compartments; similar scaling is observed for static aspects of heterogeneity.  The capacity to separate such concerns into separate statecharts (or, for static heterogeneity, into different parameters) further allows changes in the number  or design of such statecharts (or in the types of heterogeneity) to be performed in a modular, localized fashion; this contrasts with the global changes that are required across the scope of an aggregate model as dimensions of heterogeneity or dimensions of progression are changed.

Finally, within an agent-based model, there is no requirement that statecharts be fully independent of one another; we can allow them interact in ways that are defined through localized interactions, but otherwise evolve fairly independently. Such interactions can be implemented in an elegant fashion compared to the combinations that are required for highly stratified aggregate models.

\subsection{Environment}
\label{sec:environment}
Agents are not solitudes; they are placed in an environment which situates them in some context. This context commonly situating agents in space and/or in networks with connections to other agents. A person can have their presence represented in one or more networks, for example, a family network, collegial network, social network, intravenous drug use network, or sexual network. While network connections between agents of the same type are common (e.g., between two persons), so are connections between agents of different types -- for example, networks connecting persons and community service providers, or population members and their physicians. Often connections within networks serve as conduits for interactions between the pair of connected agents. The most common mode of interaction over such connections -- and, by extension -- and networks is via message-passing where one agent sends a message to another.  Such message passing provides a very flexible and computationally elegant means of characterizing agent interactions along one or more networks.

\begin{figure}
\sidecaption
\includegraphics[width=\textwidth]{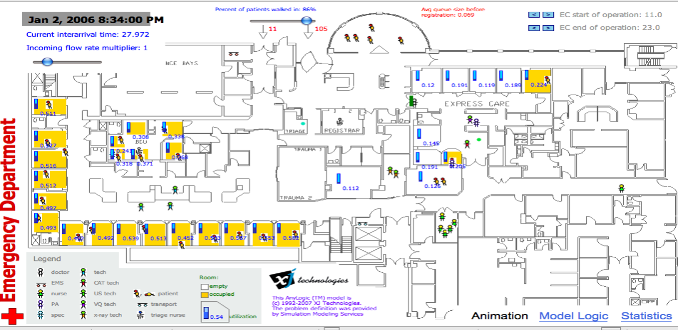}
\caption{Example of an agent-based model employing an irregular spatial environment \cite{anylogic-software}.}
\label{fig:irregular}
\end{figure}

\begin{figure}[t]
\sidecaption[t]
\includegraphics[width=\textwidth]{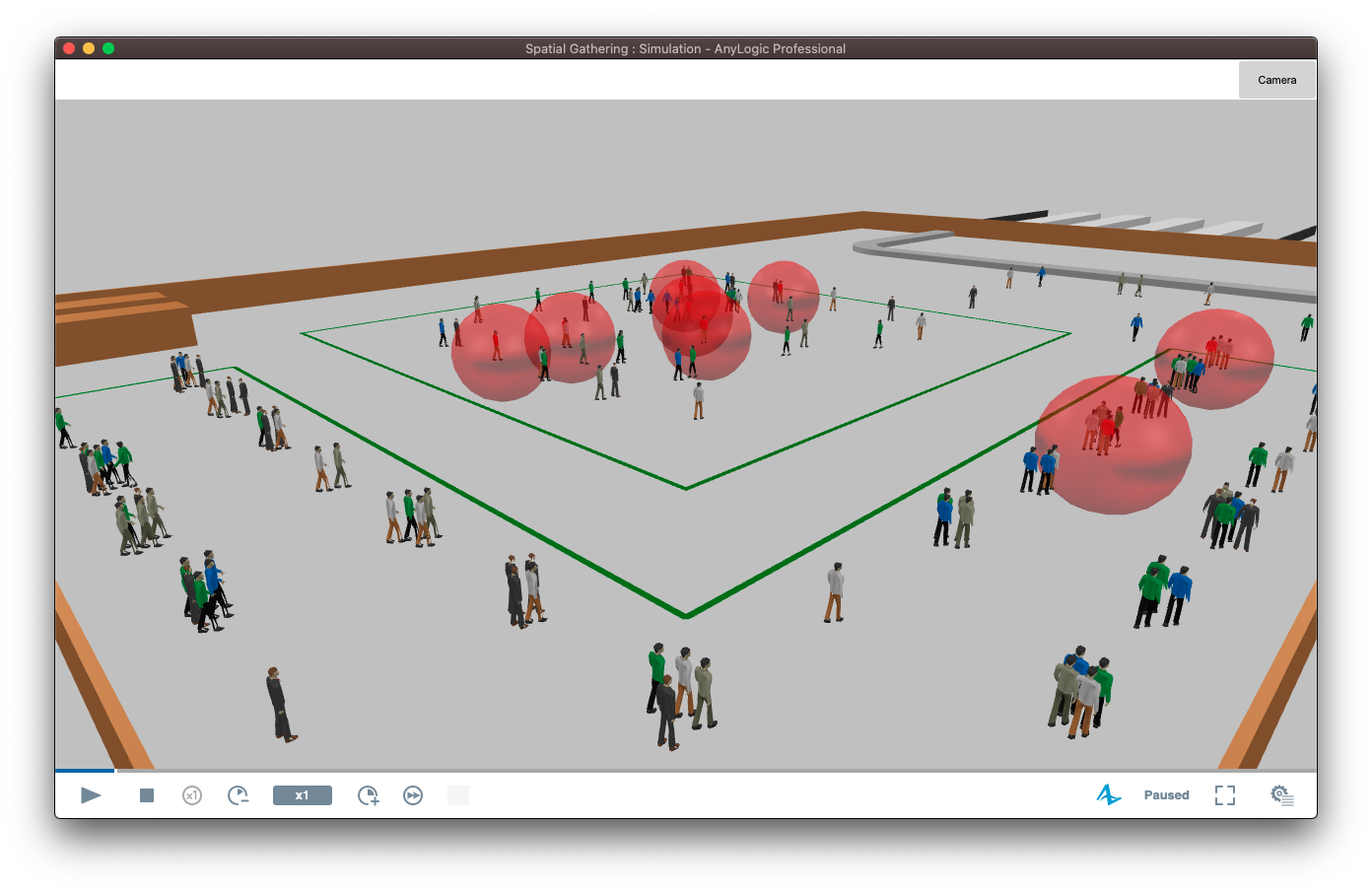}
\caption{Example of an agent-based model employing a 3D spatial environment.}
\label{fig:3dspatial}
\end{figure}

\begin{figure}
\sidecaption
\includegraphics[width=7cm]{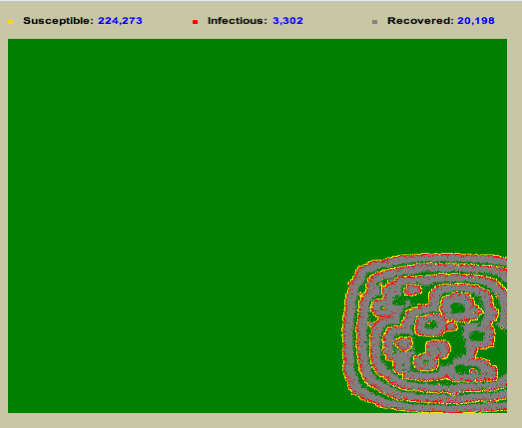}
\caption{Example of an agent-based model employing a stylized discrete square lattice spatial environment.}
\label{fig:grid}
\end{figure}

Beyond networks, we also often place agents within a spatial context. Diverse types of spatial environments can be found in the literature, including 2D or 3D Euclidean, irregular, toroidal, discrete square, triangular or hexagonal lattices, and geographic spaces. Figure~\ref{fig:irregular} shows an example of a model featuring an irregular spatial environment, Figure~\ref{fig:3dspatial} shows an example with a 3D environment, and Figure~\ref{fig:grid} shows an example with a stylized discrete square lattice spatial environment.

There are diverse motivations for placing agents in spatial environments.  Certain types of spatial environment are recommended for certain needs. With a geographic environment, we can capture effects involving locality, perception, and influence in an empirically situated spatial context. This can capture, for example, aspects of social determinants of health, like the presence of food deserts, areas with high nitrous oxide levels, or areas underserved by health care provision. Perhaps we're interested in behavior that is actually spatial in nature, like mobility changes resulting from investments in walkability or support groups for walking. Or, perhaps we're interested in disparities in COVID-19 hospitalization burden that result from the clustering of unvaccinated people or of individuals at high risk of infection due to high chronic disease burden. Situating agents in geographic context can allow us to capture the disproportionate risks in certain regions. Using models featuring spatial context allows us to look beyond an average burden  of a disease or policy and consider key local variations -- often pockets where outbreaks or policy resistance occurs where averages measures would suggest that none should occur.

\subsection{Outputs and Emergent Behaviour}
\label{sec:outputs}
A key need with all models is to understand their behaviour over time.  We often are keenly interested in emergent behavior evinced in model outputs; sometimes those emergent behaviour patterns can be quite eye-opening or surprising.
Sometimes these outputs correspond to factors that govern the evolution of model state.  In other cases, we examine epiphenomenal factors that characterize some aspect of model state (often serving as summaries of sorts), but do not drive it. Like aggregate models, commonly the behaviour of an agent-based model is considered as a function of time. However, agent-based models offer a wider repertoire of behaviours. For ABMs may output measures of state or behaviour over networks or space -- for example, reporting where infection or risk concentrate in particular regions of the network or in a geographic region. Sometimes, seeing such patterns can offer great value in understanding model behaviour. For example, within a Chronic Wasting Disease model \cite{mejia2017social}, the concentration of risk of exposure to prions in areas near the water margin may shape infection risk in ways that manifest in unexpected impacts on the population over time.

\subsection{Stochastics}
\label{sec:stochastics}
Whilst aggregate models routinely abstract away from particular events, and focus instead on broad patterns \cite{richardson2020core, Richardson2001}, most ABMs deal with individual-level events, such as exposure to infection, recovery, events associated with care-seeking, or decisions as to where to seek out food.  When depicting factors at the level of individual events, it is common that health ABMs consider elements of human behavior and psychology that -- given suitable model scope -- are best described as stochastic. Treating such behaviours as subject to stochastic evolution does confer advantages, such the ability to allow us to explain variability we see in real-world data, but also places an onus on us as modelers to ensure that observed model results are not merely flukes resulting from one chance event or series of events in a model but instead reflect regularities and structure within the behaviour of the system being modeled. Critical to offering well-founded scientific insights, we must ensure that results are replicable.  In order to achieve this, we typically run a model many times over and examine results from those many realizations, called an \emph{ensemble}.

\begin{figure}
\sidecaption
\includegraphics[width=\textwidth]{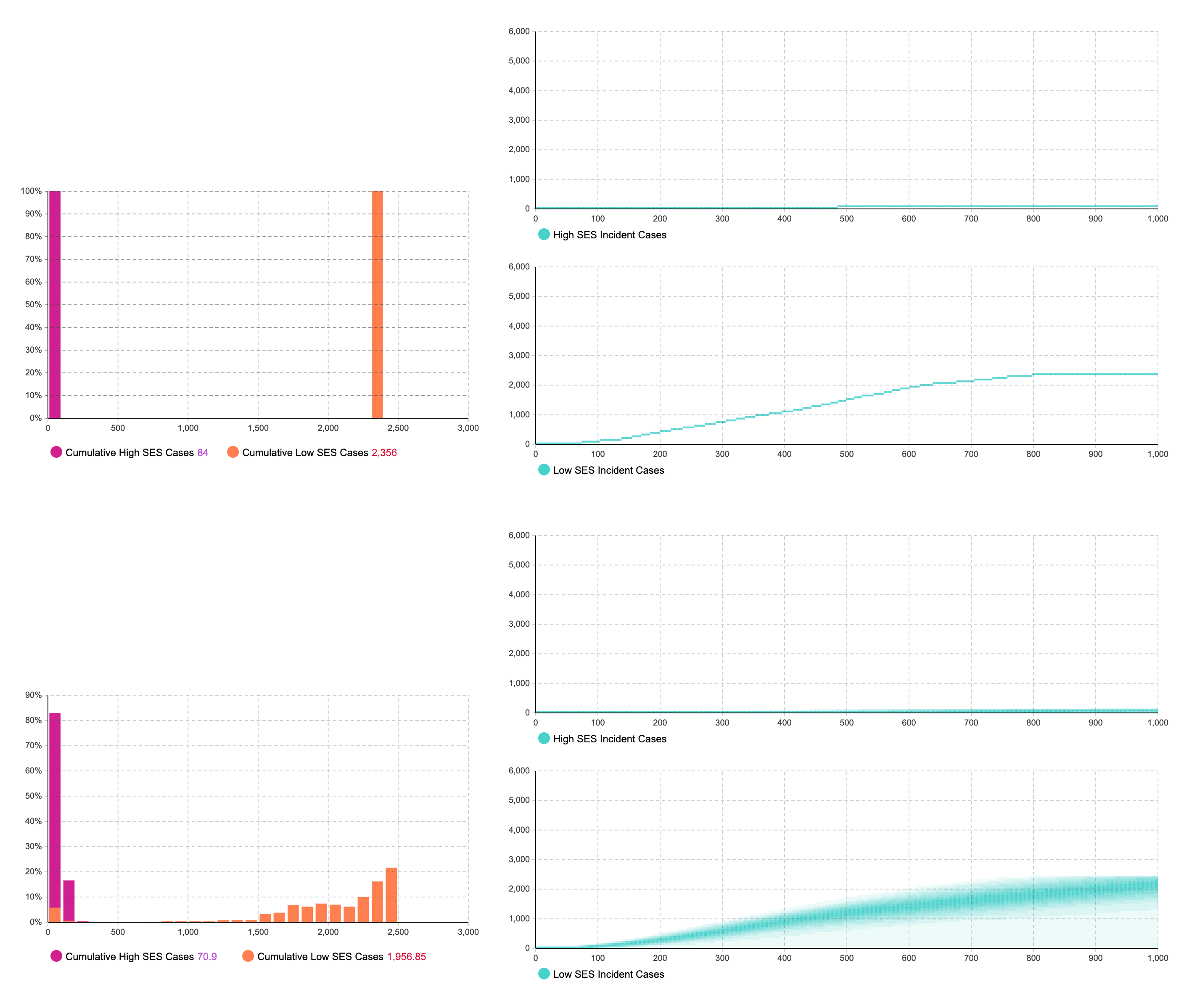}
\caption{Example of output from a single run of an agent-based model compared with that from an ensemble.}
\label{fig:ensemble}
\end{figure}

If we consider the example in Figure~\ref{fig:ensemble}, we can see the difference between a single realization of a stochastic model and an ensemble. A single run of this model reveals a marked difference in cases between the two socioeconomic status (SES) groups, with a higher case count for the low-SES group. If we consider an ensemble of many runs, it is revealed that the initial single run represents just one of many possible outcomes. The ensemble reveals that randomness in the initial state together with stochastics in the model induce empirical distributions on the case counts for each of the low- and high-SES groups. Over time, we see some realizations with smaller numbers of cases, some with larger, and some where population-wide spread never takes off in either the high- or low-SES group.

\subsection{Interventions}
\label{sec:interventions}
Many -- but not all -- of the motivation to use ABMs focuses on ``what-if" questions, reflecting a desire to evaluate interventions and counterfactuals.  Within such scenarios, we seek to ask the model `Given the assumptions of this model, what would be the impacts of undertaking action \textit{X}?' One of the most powerful things about ABMs is that they can be used to examine the effects of intervention mechanisms exploiting \textit{individual-level longitudinal information} -- with the way in which the intervention acts upon a given person depending on, for example, a person's history, in addition to their static characteristics and current state. An intervention examined using an ABM could be targeted based on, for example, a person's past presentation history or episodes of care -- for example, with a sexually-transmitted infection (STI) clinic or dental office offering behavioural counseling for individuals whose recent presentation history suggest risk behaviour, or individuals with recent acute hospitalizations could be targeted for regular dual COVID-19/influenza vaccination.  Beyond depending on individual history, intervention design and implementation can also exploit ABM's rich capacity to represent heterogeneity and context by basing intervention mechanisms on individual characteristics or based on geographic or network position. As an example, such models could represent and evaluate impacts of provisioning distinct care pathways to ensure culturally appropriate care for members of indigenous population.  As another example, control and prevention of STIs recognized the priority of focusing on core groups within STI networks,  is a key priority, and ABMs can readily characterize interventions triggered by or whose details depend on context at any number of different levels within the model can be very valuable. Or context-targeted interventions could be defined by focusing on addressing neighbourhoods situated in food deserts.  Other classes of interventions readily characterized in ABMs do not merely target certain contexts, but actively intervene upon them.  For example, an intervention may work by \textit{establishing} network connections between people -- building social capital or supportive or pro-social influences -- as some of the work of Alan Shiell and Penny Hawe \cite{ShiellHaweMaternalHealth, ShiellHaweMaternalHealthClusterRandomized} and others have investigated. In fact, many interventions focus at a certain level on changing networks --- for example, those employing support groups or community-based support organization, accountability partners, ``buddy'' systems.

\section{Example -- Chickenpox}
\label{sec:chickenpox}
The first example model characterizes the dynamics of chickenpox and shingles. This model has been employed to investigate several questions surrounding those diseases \cite{rafferty2018evaluating, rafferty2021economic, rafferty2020seeking}.

\subsection{Chickenpox and Shingles}
\label{sec:cp_epi}
 Chickenpox and shingles are two distinct diseases both caused by the varicella zoster virus (VZV). Chickenpox is typically a childhood disease; once a person has recovered from this disease, the virus generally remains dormant in the body. VZV can reactivate later in life as shingles, which causes an often debilitating and painful rash suffered by middle- to senior-aged adults \cite{campbell2010varicella, cohen2013herpes}. The first question that we investigated when building this ABM was whether vaccination for chickenpox would cause an increase in shingles incidence?

\subsection{Model Scope}
\label{sec:cp_scope}
It is often helpful to characterize model scope by describing which features are endogenous to the model, which are exogenous and specified by the modelers, and which are ignored \cite{BusinessDynamics}. Endogenous factors in this model include transmission, contact patterns, mother-child dyads, vaccination schedule adherence, fertility, mortality, hospitalizations due to VZV infections, waning of disease-induced immunity, boosting of immunity, and accumulated costs. Implementation of endogenous mechanisms are discussed further in Section \ref{sec:cp_statecharts}. Exogenous factors included assumptions about vaccine attitude, the specified vaccine schedule, and assumptions about the effectiveness of the vaccine, which were drawn from clinical research. Additional exogenous factors included initial population demographics as well as population density and unit costs assumed in the various cost effectiveness analyses. Ignored factors include household structure, time variation of contact structures, schools, and child care facilities. Key parameter values for the Chickenpox model are listed in Table \ref{tab:cp_parameters}.

\begin{table}[!t]
\caption{Key Parameter Values for Chickenpox Model}
\label{tab:cp_parameters}       
%
%
\begin{tabular}{p{0.4\textwidth}p{0.4\textwidth}p{0.17\textwidth}}
\hline\noalign{\smallskip}
Parameter & Value(s) & Source  \\
\noalign{\smallskip}\svhline\noalign{\smallskip}
Initial Population & 500,000  & \\
Mortality and Fertility & Multiple & \cite{canada2008census,canada2016census}\\
Initial Cell Mediated Immunity & \texttt{max(0.001,normal(0.05,1))}  & \cite{ogunjimi2015integrating}\\
Force of Reactivation & \texttt{gamma(2,0.1,0)}  & \cite{ogunjimi2015integrating}\\
Waning of Immunity Coeff. Shingles & 0.45--0.93 & Calibration\\
Waning of Immunity Rate Shingles & 0.4 year$^{-1}$ & \cite{ogunjimi2015integrating}\\
Duration of Exogenous Boosting & 0.42--10.0 year & Calibration\\
Exogenous Infection Rate & 17.83 year$^{-1}$ & Calibration\\
Prob. of Inf. on Contact (Normal) & 0.78 & Calibration\\
Prob. of Inf. on Contact (Breakthrough) & 0.234 & \cite{takahashi2012varicella}\\
Prob. of Inf. on Contact (Shingles) & 0.234 & Calibration, \cite{takahashi2012varicella}\\
Connection Range Normal & 8.958 & Calibration\\
Connection Range Preferential & 21.245 & Calibration, \cite{mossong2008social}\\
Preferential Contact Rate & 20 & Calibration\\
Normal Contact Rate & 30.124 & Calibration, \cite{mossong2008social}\\
Shingles Connection Range Modifier & 0.124 & Calibration\\
Preferential Mixing Age & 1--9 years & \cite{kwong2008impact}\\
Population Density Urban & 0.3 & \\
Population Density Rural & 0.2 & \\
Vaccination Attitude & Acceptor = 65\%, Hesitant = 30\%, Rejecter = 5\% & \cite{alberta2017IHDA}\\
Probability of Catch-Up & 55\% &  \\
Prob. Administered First Dose & Acceptor = 97\%, Hesitant = 30\%, Rejecter = 5\% & \\
Prob. Administered Second Dose & Acceptor = 98\%, Hesitant = 82\%, Rejecter = 33\% & \\
Primary Vaccine Failure First Dose & 16--24\% & \cite{takahashi2012varicella,bonanni2013primary,duncan2017varicella}\\
Primary Vaccine Failure Second Dose & 5--16\% & \cite{takahashi2012varicella,bonanni2013primary,duncan2017varicella}\\
Waning of Vaccine Immunity 1 Dose & 0.02 year$^{-1}$ & \cite{takahashi2012varicella}\\
Waning of Vaccine Immunity 2 Dose & 0.00 year$^{-1}$  & \cite{takahashi2012varicella}\\
\noalign{\smallskip}\hline\noalign{\smallskip}
\end{tabular}
\textit{PeerJ} 6:e5012 (\texttt{https://doi.org.10.7717/peerj.5012}). CC BY 4.0 \cite{rafferty2018evaluating}
\end{table}

\subsection{Statecharts}
\label{sec:cp_statecharts}

\begin{figure}
\sidecaption
\includegraphics[width=\textwidth]{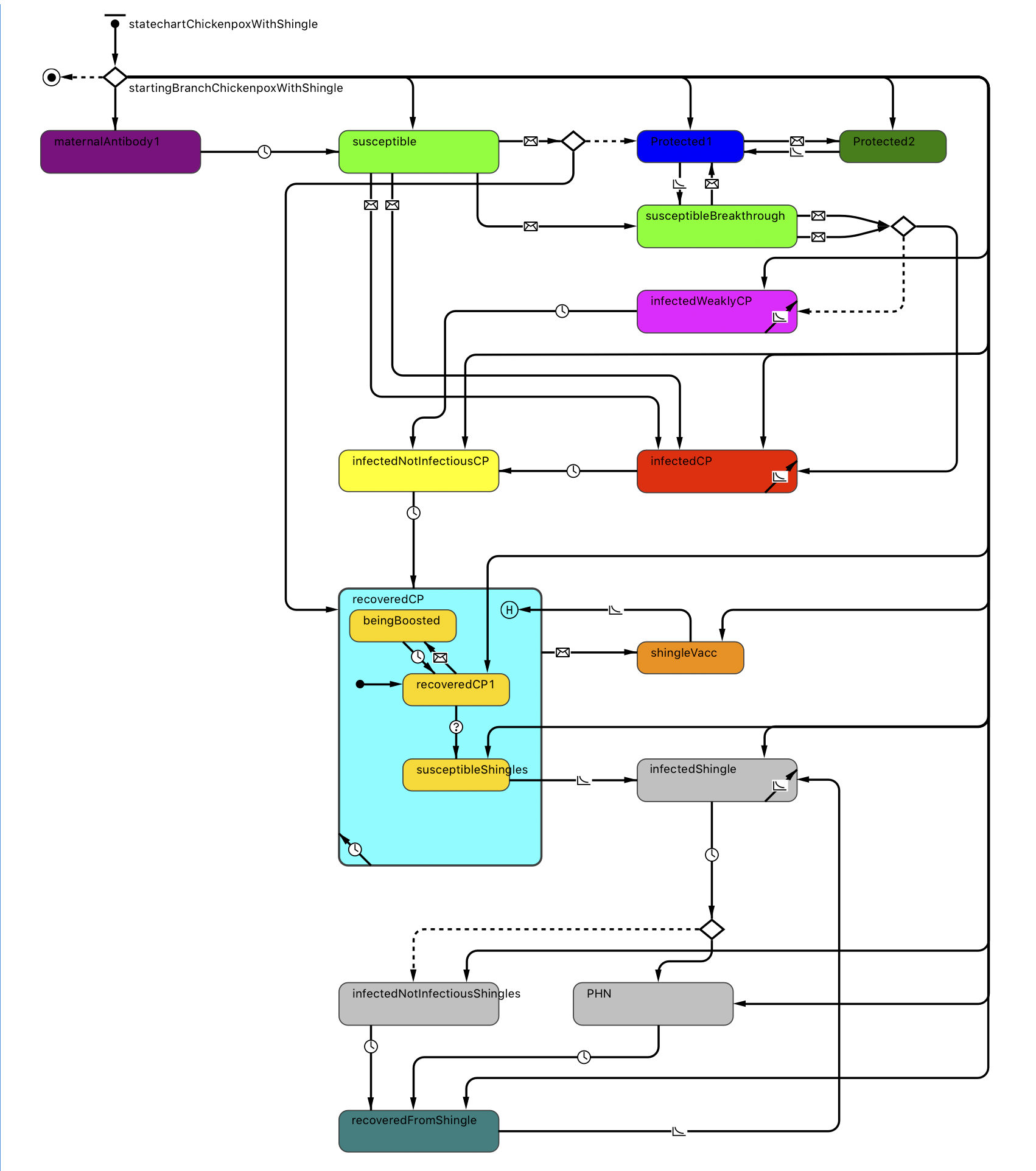}
\caption{Statechart for natural history of chickenpox and shingles. Image from Rafferty et al. (2018) Evaluation of the effect of chickenpox vaccination on shingles epidemiology using agent-based modeling. \textit{PeerJ} 6:e5012 (\texttt{https://doi.org.10.7717/peerj.5012}). CC BY 4.0 \cite{rafferty2018evaluating}}
\label{fig:cp_statechart1}
\end{figure}

The statechart in Figure \ref{fig:cp_statechart1} represents the natural history of disease for the varicella zoster virus. Following a child's birth, they are protected for some months by maternal antibodies; they then become susceptible. Two states represent protection due to vaccination; with only one dose it is possible to have a breakthrough infection while vaccinated, but once two doses are administered a person is considered to be immune for life. A person may go on to be weakly or fully infected with chickenpox. During the ensuing period of infectiousness they may transmit the infection to others through exposure messages they send (according to Poisson arrivals) to nearby agents; there is then a period where they continue to show symptoms but are no longer infectious to others. Once a person recovers from chickenpox, they enter the recoveredCP state, within which there is a mechanism that represents a process of episodic boosting of immunity driven by exposure to other people with chickenpox or shingles. Although it has not yet been required in investigating any of our scientific questions that we addressed with the model, the model further characterizes the occurrence of shingles vaccination. The remaining states represent progression of shingles infection, which can be present as a mild case or a severe case, with the latter being designated as post-herpetic neuralgia (PHN) in the statechart. A person afflicted by shingles will eventually recover, after which they are subject to a certain chance of relapsing.

\begin{figure}
\sidecaption
\includegraphics[width=7cm]{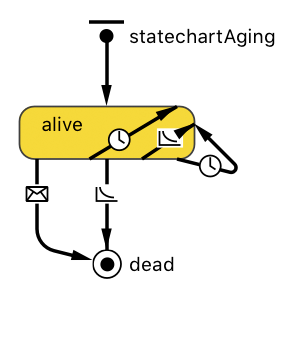}
\caption{Statechart for life and death in the chickenpox model. Image from Rafferty et al. (2018) Evaluation of the effect of chickenpox vaccination on shingles epidemiology using agent-based modeling. \textit{PeerJ} 6:e5012 (\texttt{https://doi.org.10. 7717/peerj.5012}). CC BY 4.0 \cite{rafferty2018evaluating}}
\label{fig:cp_statechart2}
\end{figure}

Life and death are handled by another state chart in this model, shown in Figure \ref{fig:cp_statechart2}. A person is alive for the duration of their lifetime, with mortality occurring according to a certain background death rate or due to VZV infection (realized by receipt of a message for death sent from the infection process within the same agent). Certain events, including childbirth for females, happen during their life, as represented by the arrows within the alive state. The external self-transition updates information periodically while the person is alive.  It also bears note that the model includes natality processes, and newborn babies enter the statechart through the initial transition, which extends down from the ``statechartAging'' label.

\begin{figure}
\sidecaption
\includegraphics[width=7cm]{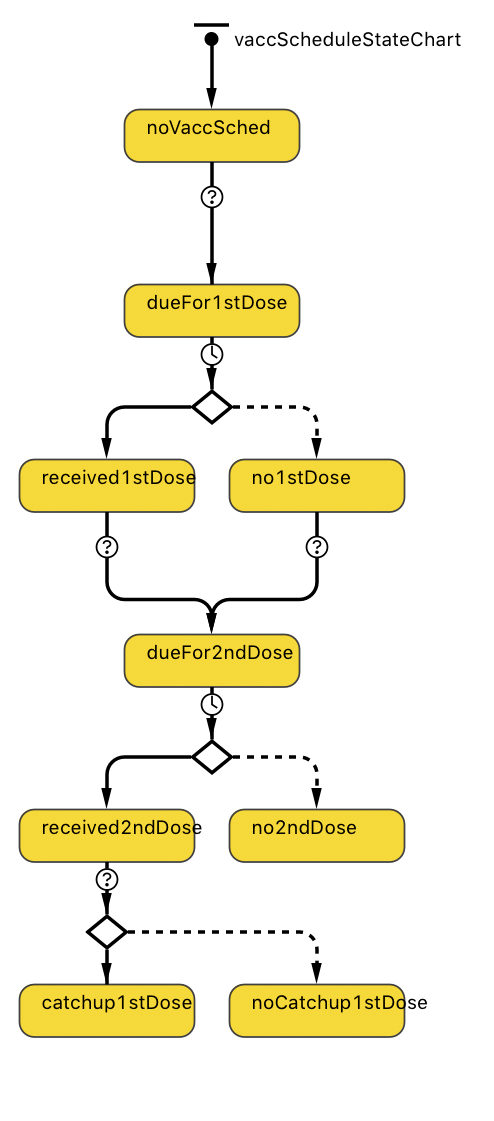}
\caption{Statechart for vaccination in the chickenpox model. Image from Rafferty et al. (2018) Evaluation of the effect of chickenpox vaccination on shingles epidemiology using agent-based modeling. \textit{PeerJ} 6:e5012 (\texttt{https://doi.org.10. 7717/peerj.5012}). CC BY 4.0 \cite{rafferty2018evaluating}}
\label{fig:cp_statechart3}
\end{figure}

A further statechart --- shown in Figure \ref{fig:cp_statechart3} --- represents a person's adherence to the vaccination schedule. When a person is too young to be vaccinated, they have no scheduled vaccines. They will subsequently become due for their first dose and either receive it or not, and eventually become due for a second dose and receive it or not. In the event they receive the second dose but missed the first, they may get a catch-up dose.

Heterogeneity is represented by these parallel statecharts; all of the statecharts described above were part of the person agent.  Within a given agent, all such statecharts operate concurrently, with each describing different aspects of that person's state and actions by which that state evolves. Age is captured as a continuous quantity in this model, without the coarse-graining common in stratified aggregate models.

\subsection{Model Fit}
\label{sec:cp_fit}

\begin{figure}
\sidecaption
\includegraphics[width=\textwidth]{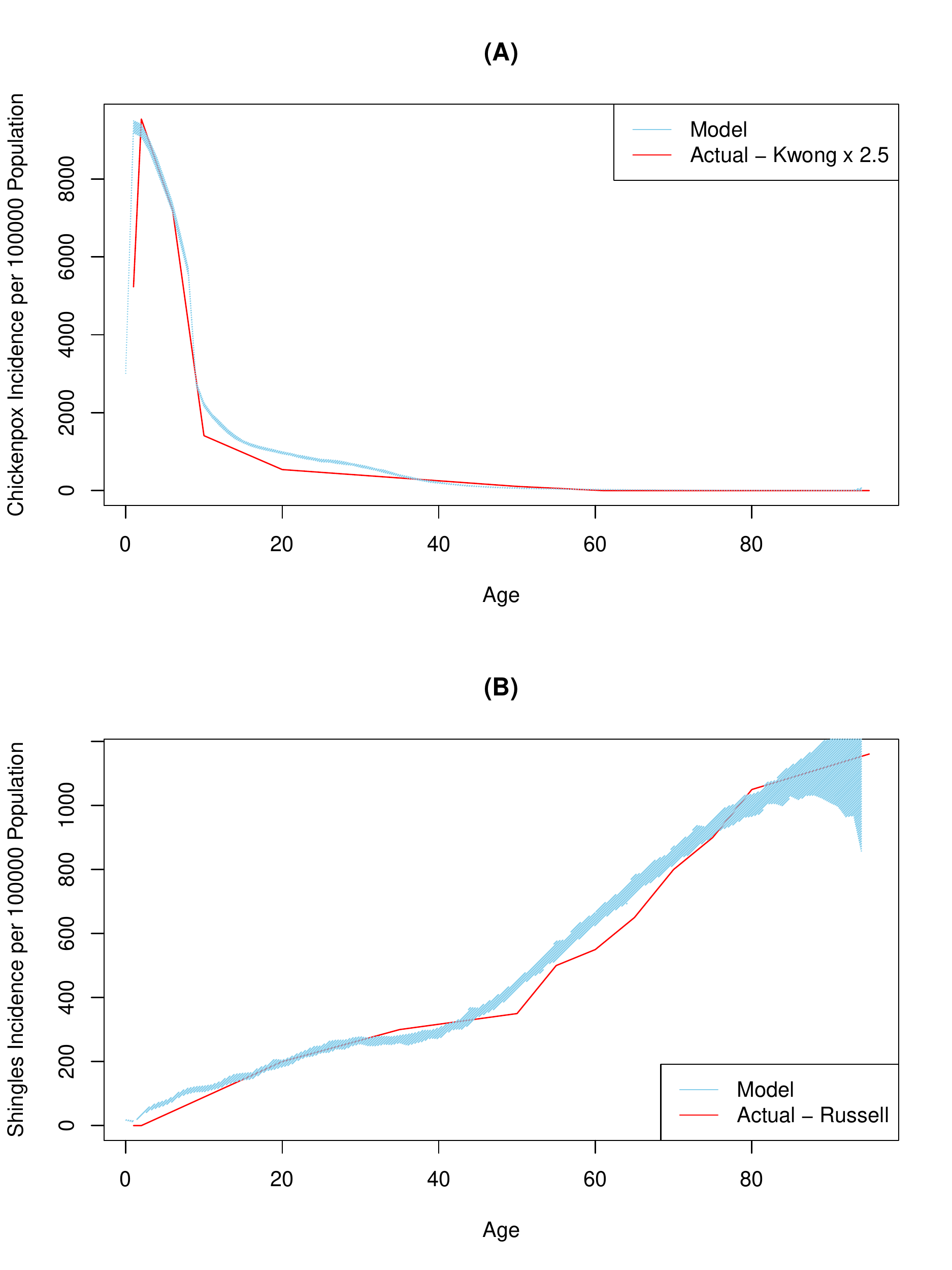}
\caption{Model fit for chickenpox and shingles incidence.  Image from Rafferty et al. (2018) Evaluation of the effect of chickenpox vaccination on shingles epidemiology using agent-based modeling. \textit{PeerJ} 6:e5012 (\texttt{https://doi.org.10.7717/peerj.5012}). CC BY 4.0 \cite{rafferty2018evaluating}}
\label{fig:cp_fit1}
\end{figure}

This model was fit to empirical data in several ways. First, we considered chickenpox incidence in the pre-vaccination era and its distribution over age groups, as seen in Figure \ref{fig:cp_fit1} (A). Red represents reference data from literature and blue represents the model output; the x-axis is age in years, and the y-axis is chickenpox incidence per 100,000 population for individuals of those ages.

A similar comparison was undertaken for shingles incidence, in Figure \ref{fig:cp_fit1} (B); again, red is is the reference and blue is the model output. The flaring of the blue at the high age point is due to progressively smaller counts of people of in oldest ages.

\begin{figure}
\sidecaption
\includegraphics[width=\textwidth]{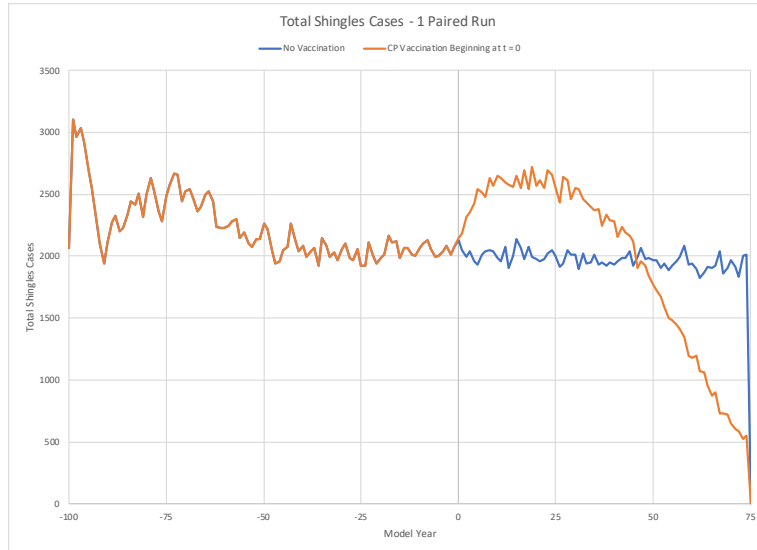}
\caption{Model result for a single paired run. \cite{rafferty2018evaluating}}
\label{fig:cp_res1}
\end{figure}

Answering the question that was posed initially, the emergent behavior that we observed in this model suggests that chickenpox vaccination is expected to cause an increase in shingles cases before it leads to an eventual decrease. Referring to the plot in Figure \ref{fig:cp_res1}; along the x-axis is time (measured in years relative to the completion of the burn-in period) and the y-axis is total shingles cases in the model, with chickenpox vaccination beginning at time 0. The figure represents one pair of realizations, consisting of a baseline and an corresponding intervention. Once vaccination is introduced, the paths diverge, and there is an increase in shingles cases for a period of about 30 to 35 years, followed by a steep decrease.

\begin{figure}
\sidecaption
\includegraphics[width =\textwidth]{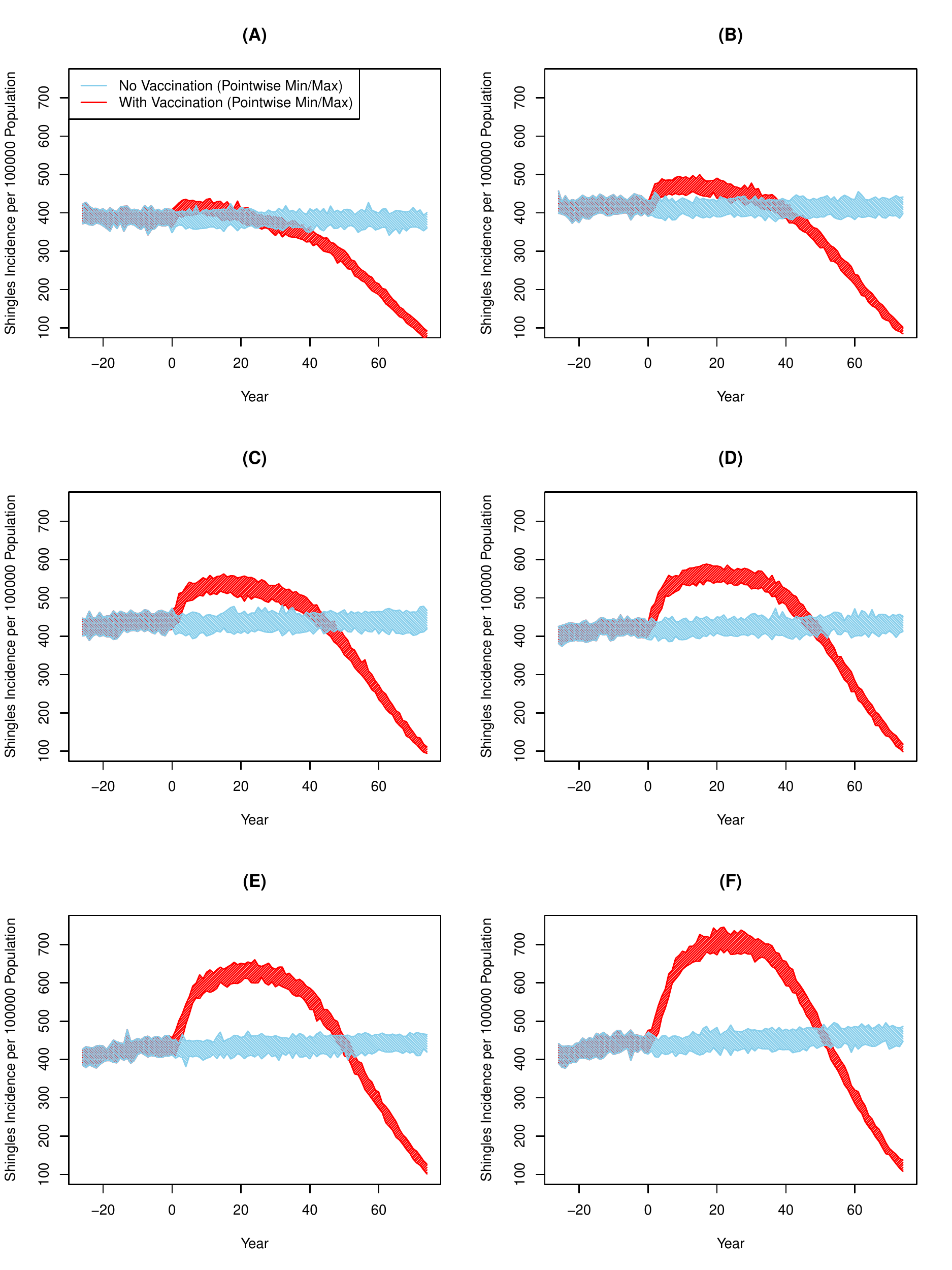}
\caption{Chickenpox model output varying duration of boosting: 2 years (A), 3 years (B), 4 years (C), 5 years (D), 6 years (E), and 7 years (F). Image from Rafferty et al. (2018) Evaluation of the effect of chickenpox vaccination on shingles epidemiology using agent-based modeling. \textit{PeerJ} 6:e5012 (\texttt{https://doi.org.10.7717/peerj.5012}). CC BY 4.0 \cite{rafferty2018evaluating}}
\label{fig:cp_res2}
\end{figure}

The degree to which VZV exposure boosts immunity remains uncertain clinically \cite{ogunjimi2013herpes}, so we conducted a sensitivity analysis on this model by examining the outcomes of assuming different durations of boosting. Figure \ref{fig:cp_res2} shows those outcomes for six alternative durations.  The panel in the upper left is the result from assuming the briefest duration of the boosting effect; the assumed durations progressively increase between panels from left to right and then top to bottom, with the bottom right panel depicting a situation where we have we assumed the longest duration of boosting.  Such results reveal another emergent behavior, which is that the degree to which shingles cases are expected to increase following the introduction of vaccination --- and the duration of time over which the number of such cases exceeds that expected absent vaccination --- depends on assumptions about exposure-induced boosting of immunity. The mechanism is a cohort effect where those infected with VZV immediately prior to the implementation of a vaccination programme are denied the exposure-induced boosting that the older cohorts received due to  low extant circulation of VZV and thus are more prone to developing shingles than the older cohorts, who were infected with the natural disease and benefited from ongoing boosting due to higher VZV circulation, as well as than the vaccinated cohorts.

\subsection{Costs and QALYs}
\label{sec:cp_econ}
This model was further used to analyze some health economic questions; to do this, some assumptions regarding costs had to be introduced exogenously. These included costs of vaccine doses, costs for general practitioner and emergency department visits due to VZV, per-day costs for people who are hospitalized, personal expenses, and productivity loss costs.

Additionally, QALYs --- which stands for quality adjusted life years and is a common measure in health economics \cite{weinstein1996cost} --- were accumulated in the model based on a person's accumulated time in various states. 

\subsection{Suitability of ABM}
\label{sec:cp_abm}
The questions investigated were: 1) Assessing the impact of chickenpox vaccination on shingles incidence \cite{rafferty2018evaluating} 2) Identifying an optimum, in terms of quality of life measures, vaccination schedule for chickenpox within Canada \cite{rafferty2020seeking} (serving as an example of a longitudinal intervention)  3) Evaluating the cost-effectiveness of chickenpox vaccination with discounting \cite{rafferty2021economic}. These analyses play to the strength of ABMs in that they would be far more cumbersome, or impossible, to investigate using an equally fine-grained lens with compartmental modeling techniques.

It is worth pointing out that other modeling studies \cite{ouwens2015impact,brisson2010modeling} have approached similar research questions using compartmental modeling techniques. In comparison with the example presented here, these works made strict assumptions about the effect of boosting and, necessarily, employed coarse-grained age groups, and were limited in their ability to examine the relationship between individual vaccination and exposure history and health outcomes. 

\subsection{Choice of AnyLogic as a Tool}
\label{sec:cp_al}
While there are many software packages that facilitate the programming of ABMs, the authors feel that AnyLogic if preferable for some of their work due to its ability to combine Agent-Based, System Dynamics, and Discrete Event Simulation logic in the construction of hybrid models. While this strength is not brought to bear in the examples presented, which are purely agent-based, this is a key factor motivating author selection of AnyLogic as the tool for a number of projects.  A further central motivator for this project was the capacity of AnyLogic's declarative modeling language to communicate model assumptions and logic to health scientists on the modeling team lacking computational training, and to allow such scientists to directly critique, refine and manipulate and modify model assumptions and scenarios.  Extending accessibility and transparency of the large majority of model assumptions across the entire interdisciplinary team can greatly reduce risk of misunderstandings, miscommunications and resulting model design and implementation errors. By facilitating more effective team science, such transparency materially elevates the team's ability to produce rigorous, relevant and impactful models.

\section{Example -- Pertussis}
\label{sec:pertussis}
The second example to be discussed is a simulation model of pertussis \cite{hempel2022evaluating} which was also developed using the agent-based methods with AnyLogic software.

\subsection{Pertussis}
\label{sec:per_epi}
Pertussis, commonly known as whooping cough, is a respiratory infection caused by the \textit{Bordetella pertussis} bacterium and transmitted by droplets in the air. Symptomatic individuals develop a characteristic whooping sound --- from which the common name is derived --- as they gasp for breath after extended bouts of coughing. Infants are at high risk of serious and even lethal complications; two-thirds experience trouble breathing, and half are hospitalized. Adults, by contrast, are often asymptomatic or have non-specific symptoms, which may be confused with a cold or flu. Risk of complications rises with age, smoking, and pre-existing asthma or other respiratory conditions. A high rate of waning of immunity for this disease necessitates a course of six to eight vaccinations to achieve full protection; many people don't complete this entire course of vaccination \cite{CDC2019pertussis,pertussis2019canada,tan2015pertussis,who2019pertussis}.

In Canada, the pre-vaccination era incidence of pertussis was characterized by multi-year cycles, infecting mostly children. Upon widespread vaccination, pertussis incidence plummeted. The whole-cell vaccine was the first vaccine developed and it achieved high efficacy but suffered from a higher risk of side effects compared to later-developed vaccines. There were issues, beginning as early as the 1970s and 80s, with misinformation instigating vaccine hesitancy and impairing vaccine coverage \cite{pertussis2019canada, smith2014pertussis}.

An acellular vaccine was introduced in the 1990s, and later the multivalent DTaP vaccine, which reduced side effects, at the cost of lower efficacy. Vaccination complacency and hesitancy caused vaccination rates to flag, and Canadian outbreaks began in earnest in the 2010s, concentrating in under-vaccinated adolescents. Of particular concern was the growing risk that such transmission imposed on infants, who constitute the primary risk group \cite{pertussis2019canada, smith2014pertussis}.

In Alberta, which was the focus of our study, vaccination rates for doses 4 and later are in the 70 to 80 percent range overall.  Critically, however, there are notable disparities in vaccination rates between families, schools, and communities, leading to poorly vaccinated geographic areas and regions of contact networks at high risk of outbreaks. Notable outbreaks have occurred in the last 10 to 15 years, many in under-vaccinated adolescents who are clustered in communities with a high density of vaccine-hesitant or vaccine-refusing individuals \cite{liu2017epidemiology}.

\subsection{Model Scope}
\label{sec:per_scope}
Before characterizing model structure, a few words about model scope are in order. Endogenously represented factors within the model included transmission of pertussis, contact patterns, schools and school transitions, household structure, household formation once a person reaches adulthood, immune protection --- represented as active and passive protection on a continuous-state basis --- family vaccination adherence, vaccine catch-up for those who miss doses but remain accepting of vaccines, fertility, mortality, differential case reporting based on infection severity, vaccine effectiveness, and the population preventable fraction, a measure related to vaccine effectiveness. Exogenously specified elements included vaccine attitudes, demographics of the initial population.  Hyperparameters for some distributions for other parameters --- including vaccine attitude, immune memory types and waning rates where whole-cell, acellular, and natural infection-induced immunity --- were represented distinctly. School count and characteristics and age-specific ascertainment rates for more serious infections were also specified exogenously. Ignored were household type change, inter-regional mixing, pertussis hospitalizations and mortality. Other than contacts in homes and schools, the model also ignored occurrence of contacts during the day, including those at workplaces and childcare venues.

\subsection{Model Structure}
\label{sec:per_strucutre}
\begin{figure}
\sidecaption
\includegraphics[width=\textwidth]{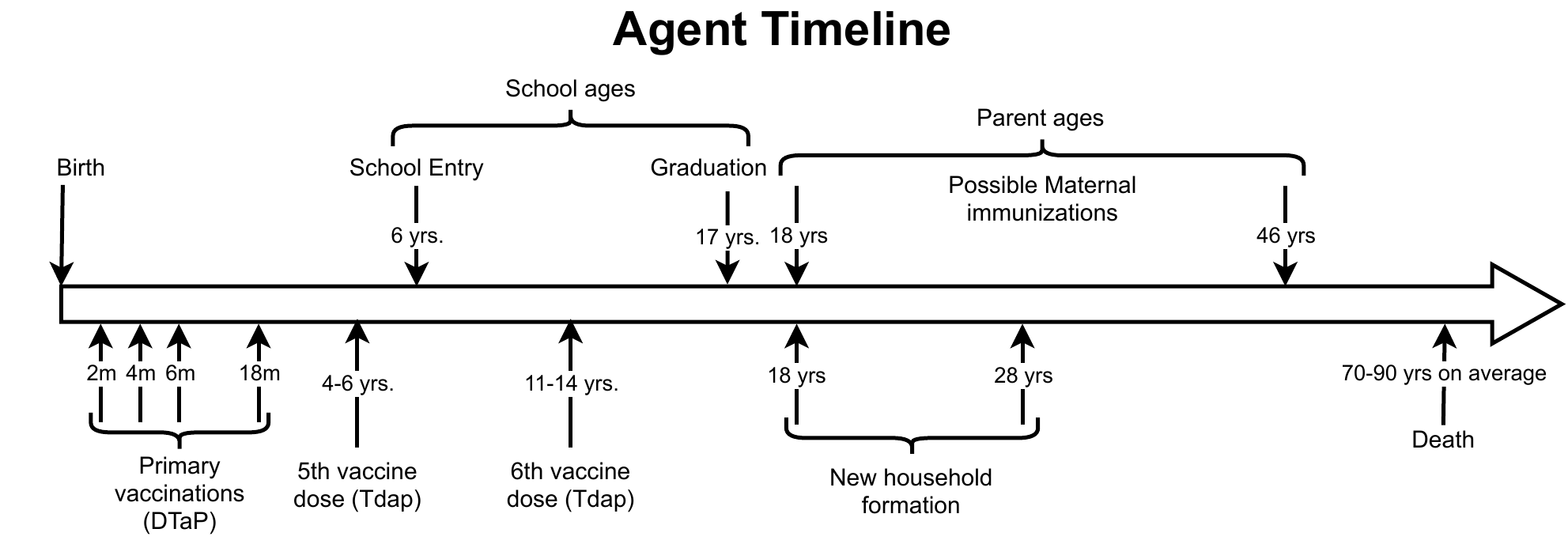}
\caption{Agent timeline for pertussis model. \cite{hempel2022evaluating}}
\label{fig:per_timeline}
\end{figure}

Figure \ref{fig:per_timeline} represents the timeline of an agent over its life course, beginning with birth. Vaccination occurs at various points mostly during childhood; children enter school at six years and complete at 18 years; between the ages of 18 and 28, children move out and form their own households and can become parents on their own up to 46 years. On average, 70 to 90 years is the life expectancy.

\begin{figure}
\sidecaption
\includegraphics[width=7cm]{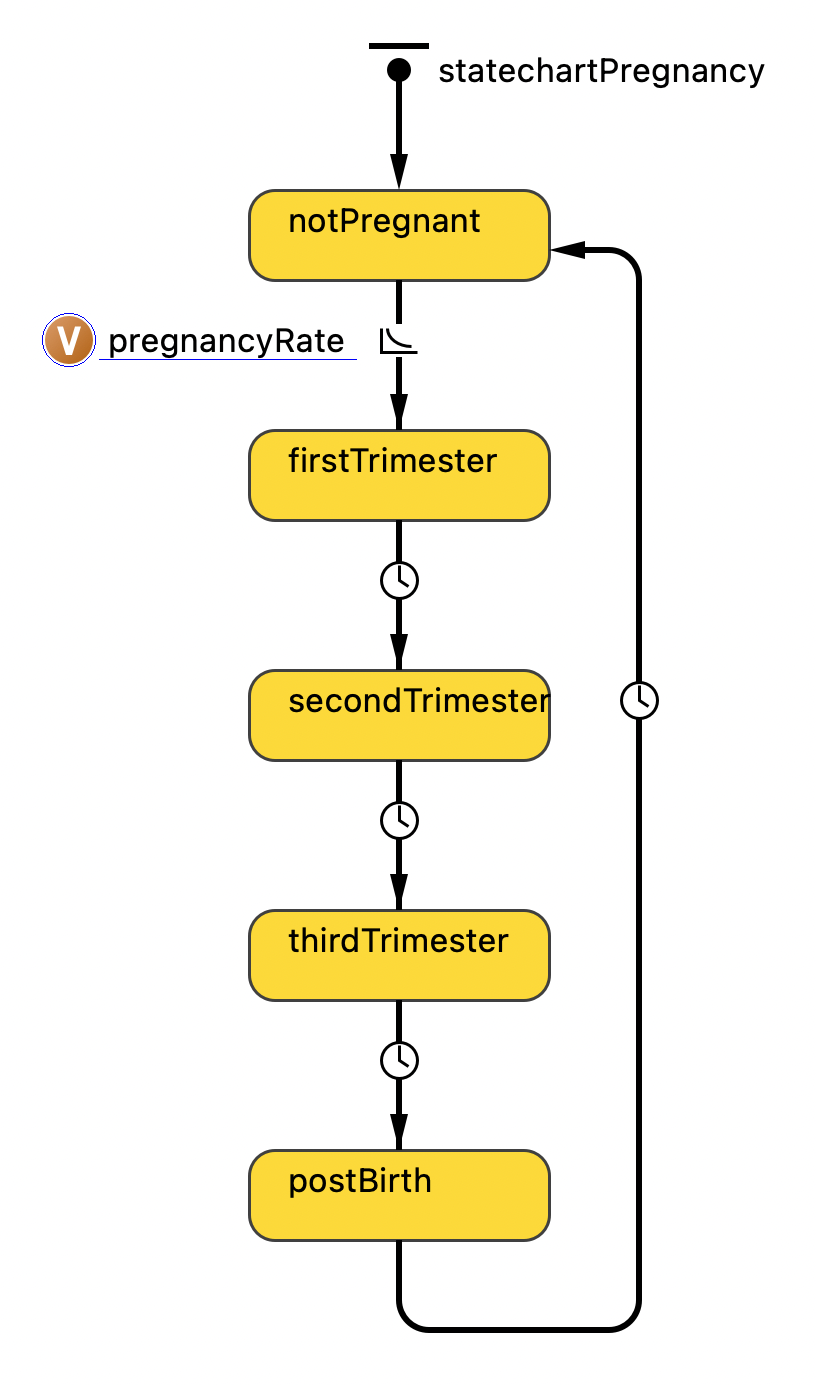}
\caption{Fertility statechart in the pertussis model.  \cite{hempel2022evaluating}}
\label{fig:per_fertility}
\end{figure}

Pregnancy and fertility are represented with an age- and family-structure-specific fertility rate. The statechart in Figure \ref{fig:per_fertility} represents states related to pregnancy, where a person begins in a non-pregnant state.  The occurrence of conception is characterized by a (hazard, i.e., temporal probability density) rate dependent on that person's fertility rate, as given by their age and the number of previous children that they've had.  Following conception, that person progresses through the trimesters, and ends in a post-partum period. 

\begin{figure}
\sidecaption
\includegraphics[width=7cm]{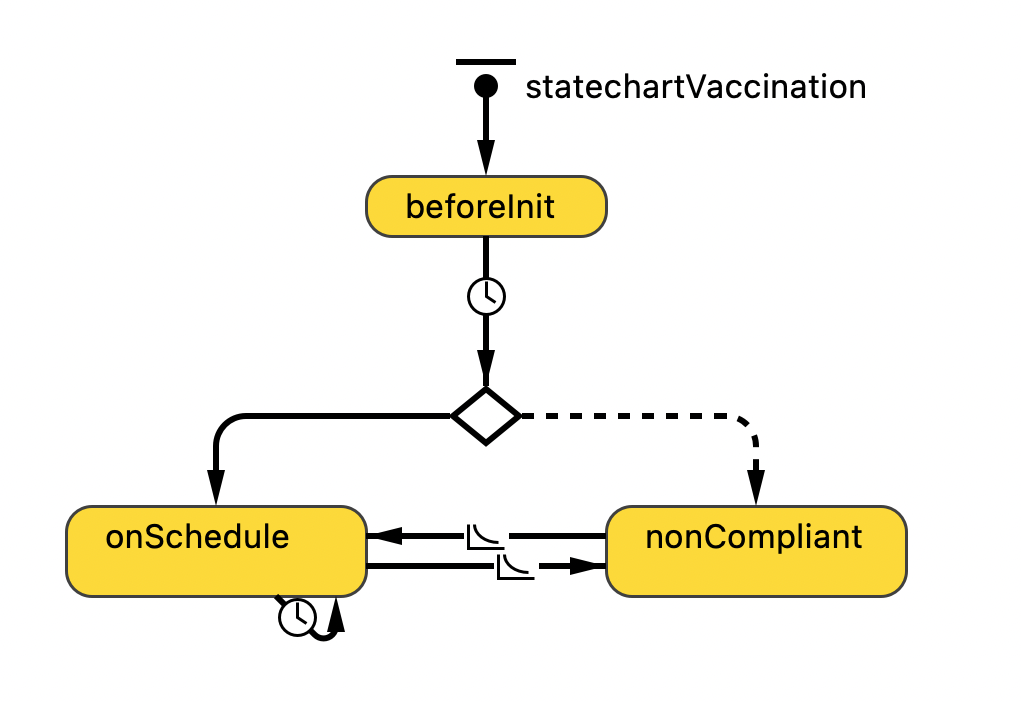}
\caption{Vaccine compliance statechart in the pertussis model.  \cite{hempel2022evaluating}}
\label{fig:per_attitude}
\end{figure}

Vaccine scheduling and compliance involved people being either \emph{on schedule} or \emph{non-compliant} and subject to a hazard rate of switching between those states at rates dependent on their vaccine attitude, as represented in the statechart shown in Figure \ref{fig:per_attitude}. Each person agent in the model is randomly assigned a number representing vaccine acceptance, where 1 represents full acceptance and 0 full refusal. Multi-person households act according to the minimum vaccine acceptance of the parents. Distributions were calibrated to ensure that the emergent vaccine coverage from the model matched empirical data.

\begin{figure}
\sidecaption
\includegraphics[width=\textwidth]{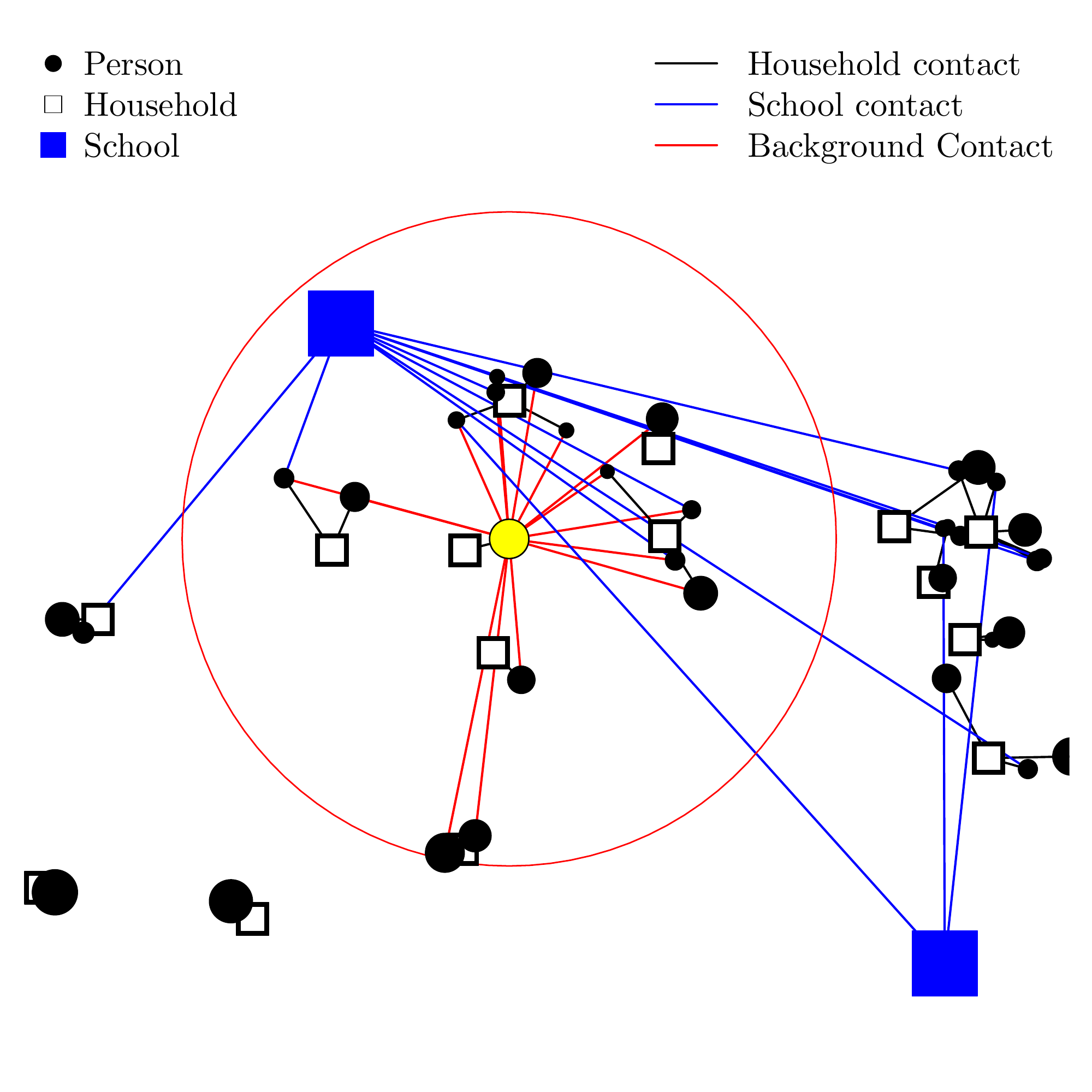}
\caption{Contact network in the pertussis model.  \cite{hempel2022evaluating}}
\label{fig:per_contacts}
\end{figure}

Within the ABM, contact patterns constitute an emergent property of the model.  Figure \ref{fig:per_contacts} depicts an example such contact network for a given person, represented by the yellow dot in this illustration; from the picture, it can be readily discerned that the network is composed of several distinct types of connections.  That index person shown in yellow is connected to all other people within a certain radius according to background contacts indicated in red. If that person is a child, they would be connected to all other children at their school. Every person is associated with a household, and is connected to all other people within their own household.

\begin{figure}
\sidecaption
\includegraphics[width=\textwidth]{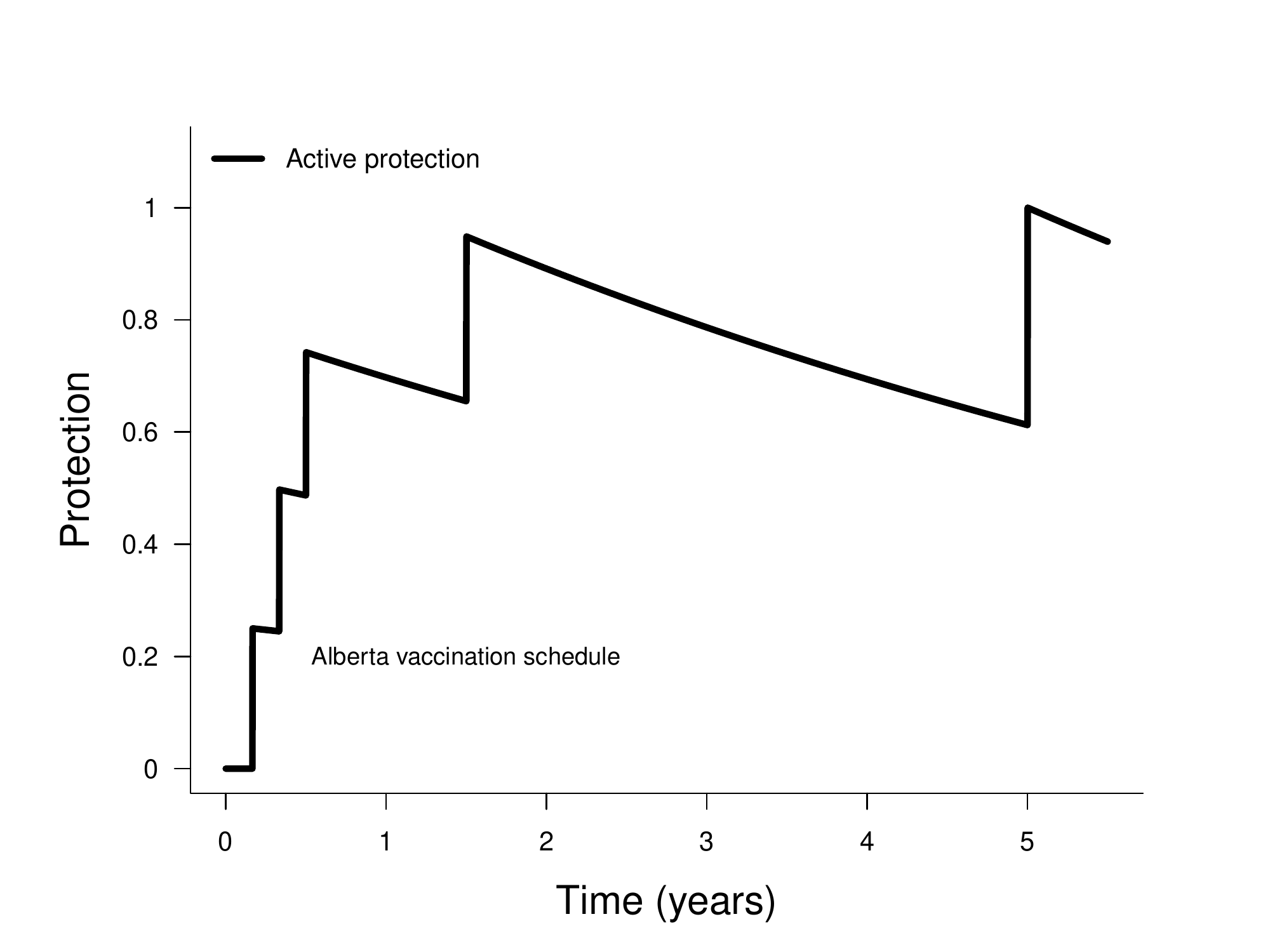}
\caption{Active protection generated by the first five doses of the vaccine course.  \cite{hempel2022evaluating}}
\label{fig:per_protection}
\end{figure}

Immunity is represented as a continuous quantity, between 0 and 1, for each person. Figure \ref{fig:per_protection} represents the development of active protection from the first five doses of the vaccine course, following the Alberta schedule for pertussis vaccination. A person begins immune naive; progression in the vertical direction is triggered by administration of a vaccine dose; between doses and absent occurrence of infection, immunity wanes exponentially.

Protection level was determined by Equation \ref{eqn:per_protection}, where maternal immunity, called passive protection, decays at a rapid rate while active protection is determined by immune memory and decays according to a rate specific to the supporting immune memory type: natural disease, whole-cell vaccine, or acellular vaccine.

\begin{equation}
\label{eqn:per_protection}
p = \min\left(p_{\mathrm{active}} + p_{\mathrm{passive}},\;1.0\right)
\end{equation}

\begin{figure}
\sidecaption
\includegraphics[width=\textwidth]{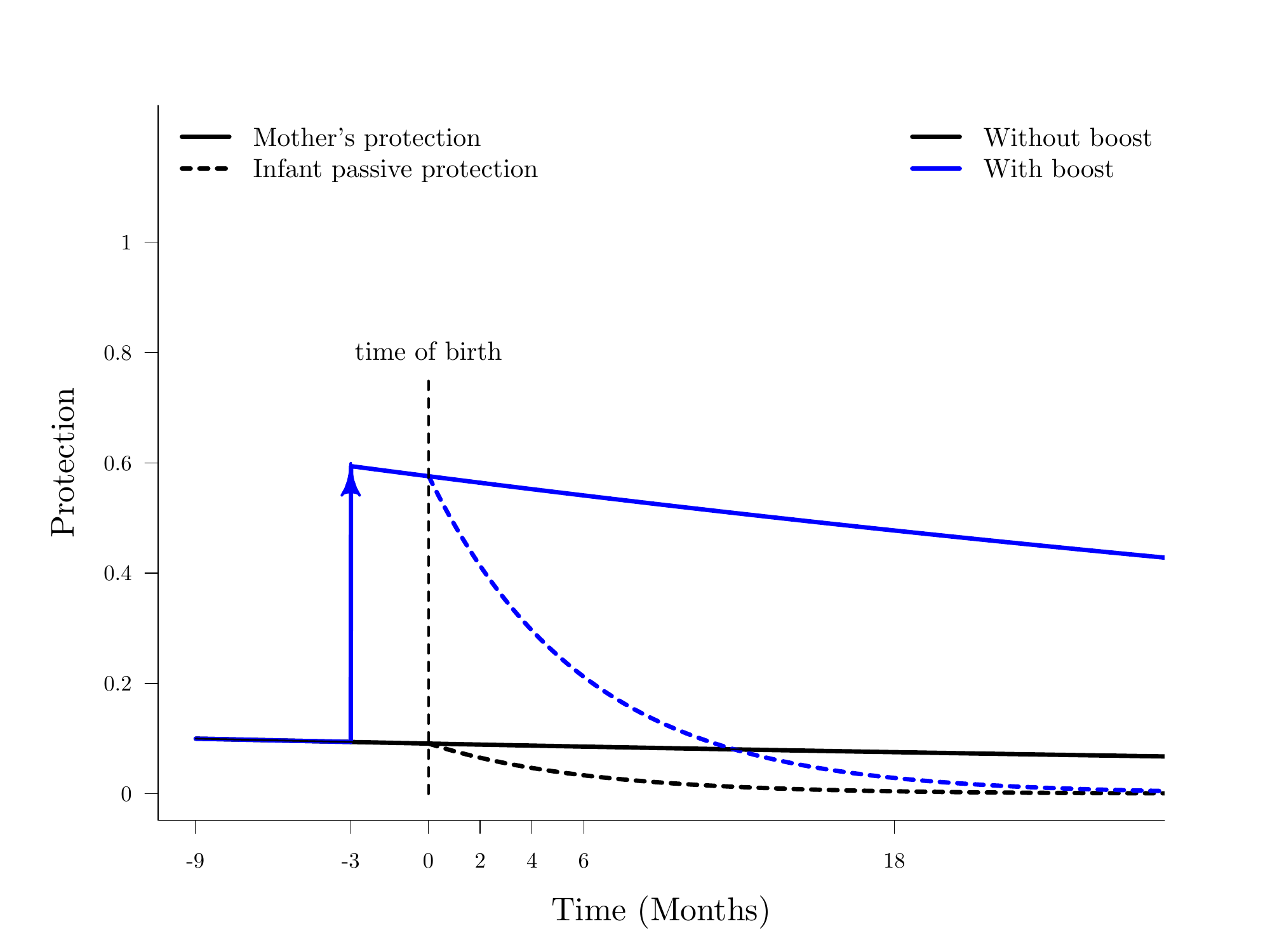}
\caption{Effect of maternal immunization on infant immunity.  \cite{hempel2022evaluating}}
\label{fig:per_maternal}
\end{figure}

Figure \ref{fig:per_maternal} represents the effect of maternal vaccination. Solid black represents the mother's immunity \emph{without} maternal vaccination and the dotted black line represents the child's immunity, which decays rapidly following birth. Solid blue represents the effect of a mother receiving a vaccination during the third trimester of pregnancy (depicted as time -3 months) and dotted blue line represents the infant under this scenario. Identical decay immunity rates obtain in both the vaccination and non-vaccination cases, but the occurrence of vaccination allows immune memory in both mother and child to decay from a higher point, leaving a higher protection level in the child for the first year of life.

A person in the model is exposed to pertussis with a certain per-exposure probability of infection $\pi$, carried in a message passed from an infective to a susceptible person. An exposed susceptible is infected with a probability $\pi$ if their protection level lies below some threshold, $\alpha_i$. Following infection, active protection is boosted to a maximum level and subsequently begins to decay. Vaccination leads to active protection being boosted incrementally as shown above in Figure \ref{fig:per_protection}. For infants, the blunting hypothesis \cite{kandeil2020immune} suggests that maternal immunization may lead to impaired immunity development by the child when receiving early doses of the vaccine.

\subsection{Model Fit}
\label{sec:per_fit}
Demographics were informed by the census population pyramid, the initial distribution of household types and distributions of couple and single households by the number of children in the household \cite{alberta2019open,canada2016census}.

\begin{figure}
\sidecaption
\includegraphics[width=7cm]{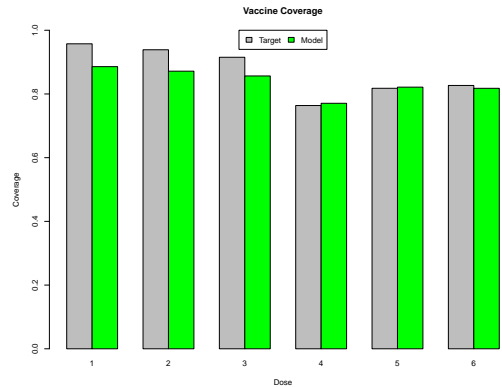}
\caption{Vaccine coverage in the pertussis model. \cite{hempel2022evaluating}}
\label{fig:per_coverage}
\end{figure}

Vaccine coverage is endogenous to the model and it was fit to actual coverage in Alberta, as shown in Figure \ref{fig:per_coverage}, which shows dose along the x-axis and vaccine coverage on the y-axis with the gray columns representing the true state in Alberta and the green columns representing what the model achieved.

\begin{figure}
\sidecaption
\includegraphics[width=7cm]{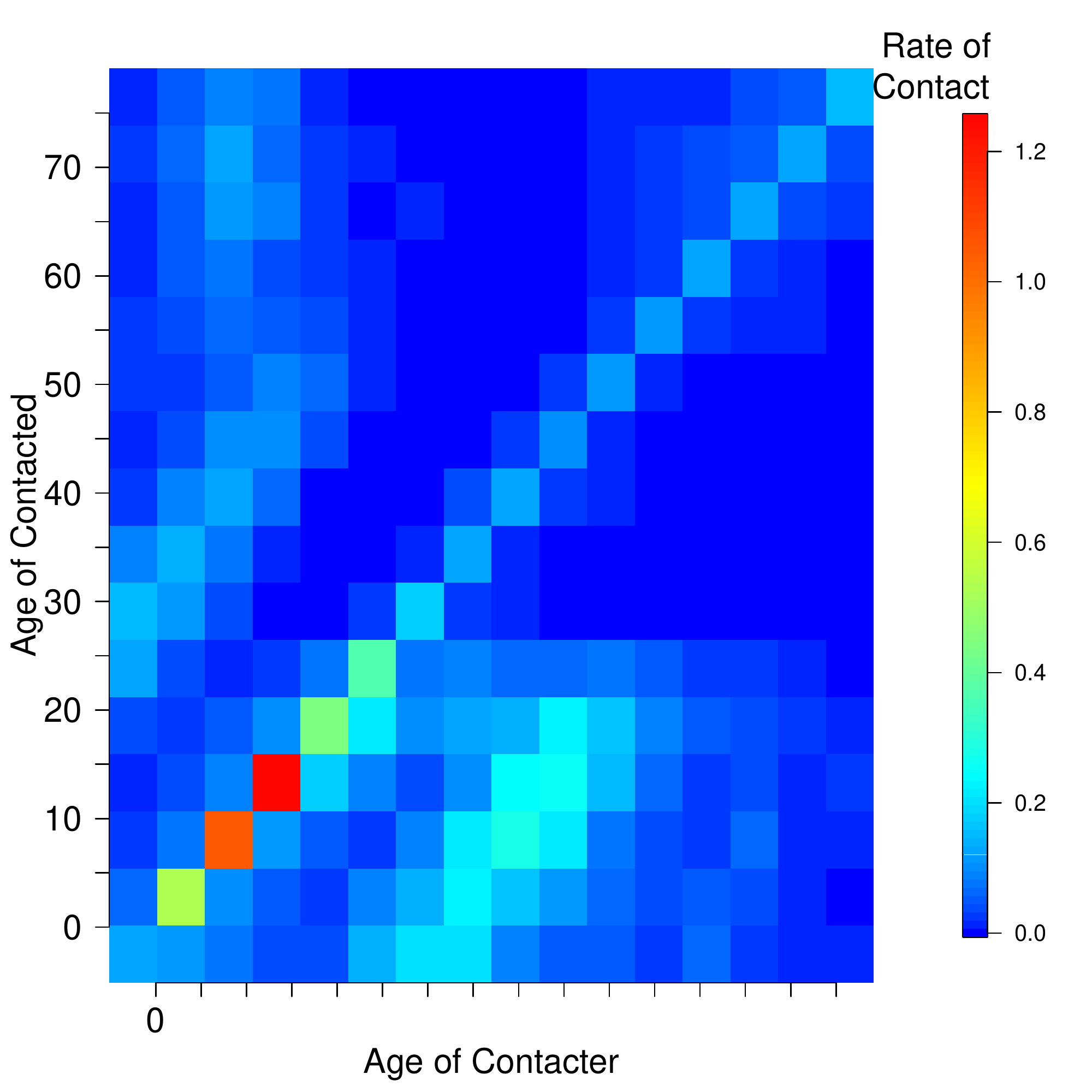}
\caption{Emergent contact patterns from the pertussis model.  \cite{hempel2022evaluating}}
\label{fig:per_contact_pat}
\end{figure}

Figure \ref{fig:per_contact_pat} represents the emergent contact patterns resulting from the contact structure described above in Figure \ref{fig:per_contacts}. The diagonal represents people contacting other people of approximately their own age, the off-diagonal areas represent child-parent contacts and then some of the other outlying areas represent child-grandparent contacts. This was compared to Mossong, et al. \cite{mossong2008social}, also known as POLYMOD, done in Europe.

\begin{figure}
\sidecaption
\caption{Pertussis model fit.  \cite{hempel2022evaluating}}
\includegraphics[width=\textwidth]{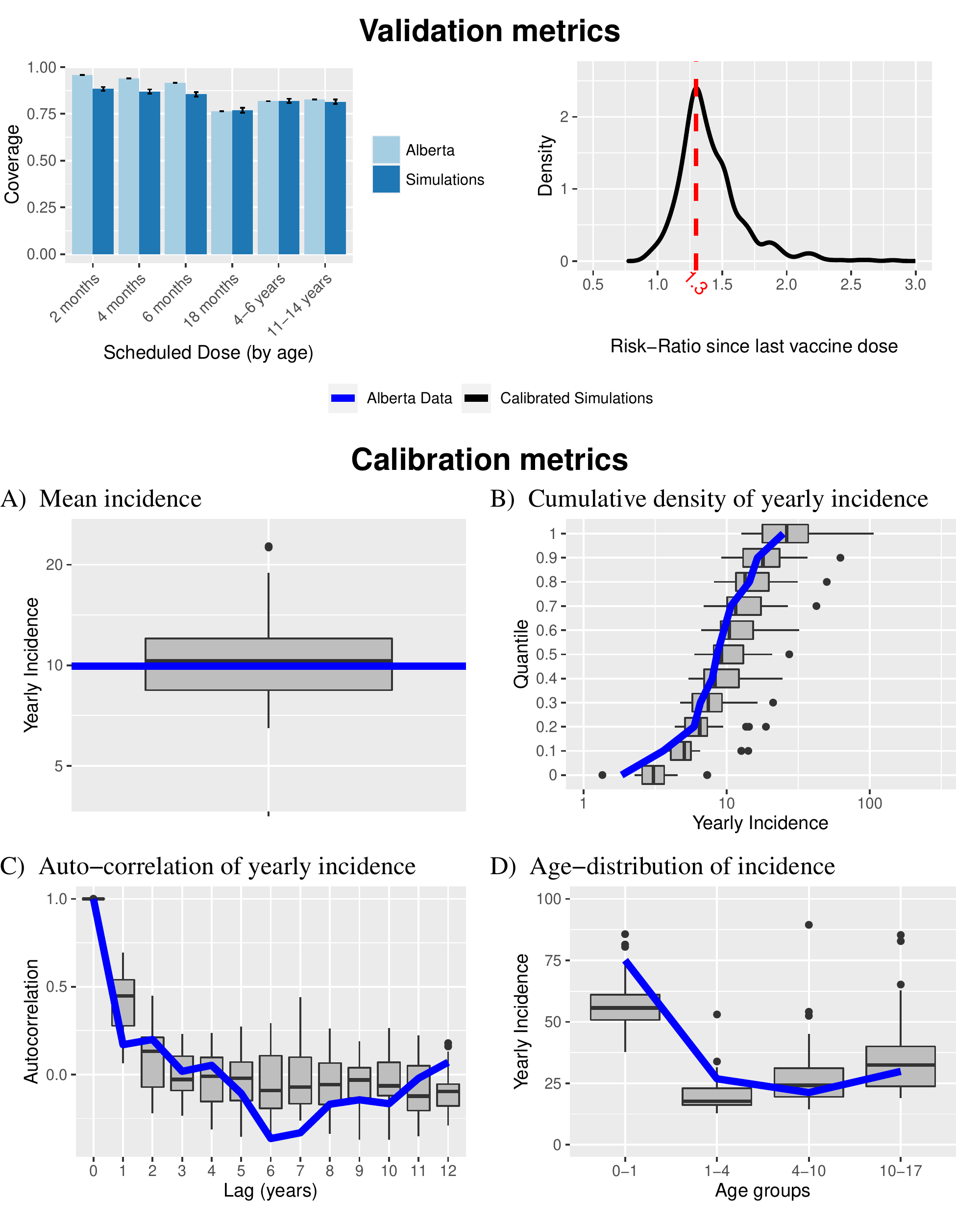}
\label{fig:per_metrics}
\end{figure}

Figure \ref{fig:per_metrics} represents calibration and validation matches; vaccine coverage was discussed above, we also matched to the average risk ratio since the last vaccination dose, mean yearly incidence, density of yearly incidence, autocorrelation of yearly incidence, and age distribution of pertussis incidence. The goal of using these measures to fit the model was to ensure that outbreaks in the model were of size and frequency that is expected based on on empirical data without trying to force it to follow every peak and valley of the actual historical outbreak pattern.

\subsection{Scenarios}
\label{sec:per_scenarios}
All scenarios involved running the model for a time horizon of 50 years with a 20-year burn-in, or warm-up, period before maternal immunization began. The initial population was 500,000 but, with an open population involving births and deaths endogenous to the model, that population expanded and shrank over time. The baseline involved no vaccination. The main intervention was maternal immunization administered across $50\%$ of pregnancies. A number of sensitivity analyses were examined, including different rates of maternal vaccination, alternative durations of maternal-antibody-based immunity, varying ascertainment rates, blunting, and no passive protection from immunization. All scenarios were assessed in ensembles of 30 realizations.

Model outcomes suggest that immunization is highly effective in reducing infant infections and substantial benefits are provided by both cocooning and transfer of passive immunity.

\subsection{Suitability of ABM}
\label{sec:per_abm}
A number of features of this study are well-suited to be examined with an ABMs. One is the need to consider network effects, particularly in light of an overall contact network emergent from a number of time-varying sub-networks. Continuous, heterogeneous state in the form of dynamics of immune memory of a person are readily captured using an ABM, but can only be discretely approximated using a stratified compartments model. As a family-level phenomenon, the adequate characterization of cocooning-based protection is difficult to achieve in a compartmental model, but readily with an ABM. Whilst readily characterized with an ABM, description of targeted intervention of immunization at a certain point during pregnancy would be highly cumbersome within a compartmental model due to the curse of stratification dimensionality. This model can also be held up as an exemplar of large compute requirements as a limitation of ABM, with a single realization taking over a day to evaluate on a dedicated computer platform.

\section{Tradeoffs Between ABMs and Aggregate Models}
\label{sec:tradeoffs}
This section briefly highlights some of the many tradeoffs between agent-based and aggregate models, including those based on both technical considerations and those associated with the modeling process and organizational context.  Those seeking further depth of discussion of such tradeoffs are referred to past contributions focusing on this subject \cite{ osgood2009Comorbidities, osgood2004Heterogeneity, osgood2007TraditionalSDAndABMToolsets, rahmandad2008heterogeneity, read2003disease, parunak1998agent, macal2009agent, bankes2002agent}.

An ABM advantage of great significance in both conceptual and practical domains lies in its ability to readily capture and nimbly evolve continuous, discrete, and relational heterogeneity, whether static or evolving over time. Straightforward and easily evolved representation of heterogeneity allows us to more readily address concerns involving health equity and disparities. We have multiple aspects of state we can readily keep track of, for example, multiple comorbid conditions or behavioral concerns.  

Whilst it represents a type of heterogeneity, the ability to capture arbitrary aspects of individual history itself is of sufficient import to merit further remark. Longitudinal information from the model can be compared against comparable empirical longitudinal information -- information that is widespread and increasingly commonly available. We can calibrate the model against such longitudinal information.  As will be further noted below, such longitudinal information can play a key role in supporting interventions. Moreover, such longitudinal information is frequently a very important class of data for understanding dynamics. 

Also of particular value to the modeling process and learning with a model is the ability to nimbly evolve the representation of heterogeneity, quickly adding new dimensions of dimensionality to the model, rapidly altering the representation of certain types of heterogeneity according to learning regarding the important distinction and measures, or removing it when judged appropriate.

Individual-based models, including ABMs, are better for examining fine-gained consequences involving network and space effects. Using multi-scale modeling, we can represent multi-level nesting of such context, representing --- in accordance with the socio-ecological model \cite{bronfenbrenner1992ecological, bronfenbrenner1977toward, bronfenbrenner1986ecology} --- the successively broad nesting of a person within a neighborhood, a school, a municipality and a country, for example. Capturing such layers of context is also advantageous to examine emergent behavior across such different levels of scale. A further advantage concerns ABM's very natural means of representing such nested contexts:  In contrast to the ``horizontal'' character of compartmental models, nesting within ABMs mirrors, in a very natural way, nesting in the world.

Although ABMs can readily handle data or data gaps that require imposing homogeneous assumptions across population members, such contexts eliminate one of the important competitive advantages of agent-based modeling.  However, the ability to scalably and nimbly characterize heterogeneity in individual characteristics is just one motivator for use of ABMs; an agent-based approach may still be recommended by its other strengths, such as the ability to capture agent history, multi-dimensional state evolution, and network and spatial effects as such factors bear on model governing processes, intervention mechanisms, or outcomes of interest.

ABM's representation of agents as situated in contexts further allows capturing of situated perception of individuals, learning over time and decision-making given those perceptions. 

Agent-based modeling's finer resolution supports characterizing and examining interventions at a far more detailed level than is readily possible with aggregate modeling. Whilst aggregate models can often be used to secure answers to coarse-grained questions as to \textit{where} to intervene in a system, ABMs can go further by examining \textit{how} to best intervene \cite{GeoffMcDonnellWhereVsHowToIntervene}. Within the sphere of interventions, ABMs can also be used to examine matters involving the \textit{implementation} of interventions, dealing with the sphere of implementation science by evaluating intervention scalability, rollout- and scale-up dynamics, financial sustainability and the time-to-effect. Also, because of ABM's ability to characterize individual progression and layers of context, we can represent interventions that are contextualized, that are based on individual position in networks or space, and we can examine interventions that are highly targeted in ways that really are not readily addressable in any plausible fashion with aggregate models. The ability of ABMs to information on arbitrary aspects of individual history capture is of such import for intervention assessment as to merit further discussion.  Such longitudinal data supports ready ABM investigation of the broad and widespread class of interventions that target or trigger interaction based on history at an individual level, or at an intermediate level of scale (e.g., at a family or neighbourhood level).  Such interventions are both common and important, and can readily involving targeting or triggering rules that are difficult to adequately characterize using aggregate models; the contrasting ease with which such interventions can be represented in ABMs renders them of great value in many public health contexts.

Of final note is ABM's capacity to support critical evaluation of broad classes of data collection and analysis methods by virtue of ABM's to characterize such methods against synthetic data in an in-silico environment in which the underlying situation (the ``ground truth'') is in fact known -- an approach commonly referred to as ``simulation experiments'' within statistics. Through such ABM-based evaluation, it is possible to more proactively identify blind spots within such data collection and analysis methods. For example \cite{OsgoodLiuPF2014} sought to use machine learning strategies such as particle filtering to assess on a recurrent basis the underlying epidemiological state of some contexts. A textured agent-based model provided a ready means of evaluating the effectiveness of such methods by serving as a source of synthetic empirical data, and comparing the particle filter-based estimates of the underlying state against the ``synthetic ground truth'' given by the actual situation within the agent-based model. The same experimental setup can ready allow for studying under what conditions the estimates are more -- or less -- accurate, the impact of network type or frequency of data collection on estimate accuracy \cite{safarishahrbijari2017predictive}.  More broadly, similar methods can be easily used -- \textit{mutatis mutandis} -- to assess the accuracy of other sampling methods, inference strategies, and adequacy of associated data collection mechanisms.

With models, it is often valuable to engage in storytelling when engaging with stakeholders -- whether people with lived experience, policymakers or other knowledge users. Models are exceptionally powerful as storytelling vehicles when we can link them up to the experiences of individuals and organizers. ABMs, in particular, can excel at this by showing or recounting as as narrative one or more simulation trajectories of individual agents (including aspects of history) or of different components of an organization \cite{YellowQuillWaterWade2022}.

All techniques have limitations, and ABMs are no exception. Explainability to non-modelers is currently a foremost challenge.  While visual depictions of dynamic models is a key asset in securing stakeholder feedback regarding and critique of such models, in the current state of the art, there is no unifying visual description language for depicting ABM structure.  Moreover, there is no widespread (much less universal) mathematical framework in which such models are specified.  Instead, the structure and rules underlying a model are operationally specified in code -- code that is almost always inaccessible to stakeholders elsewhere on model teams.  Even small ABMs commonly require a modest amount of code; medium-sized production ABMs are commonly accompanied by sizeable codebases.  Beyond impairing transparency to and critique by stakeholders, the lack of formal, transparent model specification and the frequently sprawling nature of ABM codebases pose notable problems for the communication and replication of model results that lies at the basis of scientific advances.

With all types of dynamic modeling seeking to address questions and characterize important types of factors within the world is the basic issue of model validation: `have I built the right model?'.  But with ABMs, the fundamental question of model verification --- `have I built the model right?' --- achieves particular texture, importance, and operational urgency.  This particularly reflects the fact that because of the amount of software engineering they require during the model implementation stage, ABMs often contain a great deal of programming logic where a significant number of bugs may lurk. Building and maintaining medium-sized ABMs requires not only the traditional interdisciplinary mix essential for supporting other impactful dynamic modeling projects, but also solid software engineering skills.  Such efforts place a premium on practice of quality assurance skills such as pair modeling, peer desk checks, and formal model inspections \cite{wiegers2002peer, OsgoodTianSoftwareDevelopmentDynamicModeling2012}, model testing and mocking \cite{OsgoodTianSoftwareDevelopmentDynamicModeling2012}, and continuous integration \cite{OsgoodTianSoftwareDevelopmentDynamicModeling2012}.  They also require much effort by modelers to avoid the risk of ``not being able to see the forest on account of the trees'' -- being so distracted by the welter of implementation-level software engineering detail that they lose clarity regarding and reasoning about model structure.  Finally, the large volumes of code requires interplay of skilled software engineering and savvy modeling to ensure that a model can remain capable of evolving nimbly with learning.

A key shortcoming of agent-based modeling is also the flip side of one of its key strengths:  flexibility.  Whilst that flexibility offers great advantages in crafting models that offer high-resolution lenses to investigate public health questions, it's all too easy to take the flexibility of ABMs and run afoul of it by building models where too much is included.  When building such models, it is key to apply the YAGNI principle (`you ain't gonna need it') \cite{jeffries2001extreme} and building it up in an agile fashion bit by bit \cite{OsgoodTianSoftwareDevelopmentDynamicModeling2012, abrahamsson2017agile, beck2001Agilemanifesto, kreuger2016agile}.

It's fair to say that aggregate models frequently have an edge in terms of faster (albeit more abstract) construction, lower computational burden, and greater transparency.  Because of the ability to represent aggregate models as ordinary differential equations or stochastic differential equations, they can be analyzed formally and mathematically understood in ways that are often more immediate than what is possible with ABMs. Aggregate models have lower baseline cost and involve far less programming than ABMs. In terms of run (numerical integration) time, aggregate models' computational performance costs are invariant to the population size.  The lack of stochastics in ordinary differential equation and other deterministic compartmental models means that you can run the model quite directly without as much need for ensembles. Overall, the fact that you can build aggregate models more quickly and run them more quickly leaves more time for learning and refinement. While we can say that aggregate models are often simpler, some mechanisms can be simpler to describe in ABMs, such as those many points of understanding or theory characterizing phenomena at or benefiting from description at an individual level. Representing multiple aspects of heterogeneity -- static or dynamic -- in aggregate compartmental model gives rise to a combinatorial explosion of structure. With many ABM packages, we can readily use visualization of model outputs to aid communication and intuition.  However, because of the large amounts of code involved in contemporary agent-based modeling practice, ABM structure is frequently considerably less accessible and transparent to project stakeholders compared to what is possible in compartmental modeling.

\section{Summary}
\label{summary}

Agent-based modeling, in summary, is a powerful tool for investigating health-related questions, allowing us to represent individual history, targeted interventions, whilst capturing supportive spatial, geographic, or network context. ABMs can capture agent-environment interactions in rich ways with GIS and irregular spatial networks. ABMs richly capture heterogeneity, particularly the ability to capture heterogeneity of individual history impacts, early life insults, or adverse childhood experiences, which are key for addressing specific health equity needs. Key limitations of agent-based models include computational expense: a single realization can require hours to run and requirements scale up with population; this is exacerbated by the stochastic nature of ABMs, which necessitates running an ensemble of realizations to fully capture the regularities of the model. The lack of a crisp mathematical description or visual language for ABMs impairs modeling transparency to stakeholders and communication of modeling results. For all of their tradeoffs, it is important to recognize, that recent modeling advances point us to look beyond choosing one or the other modeling method, and to the importance of judiciously weaving them together for effective hybrid modeling. We defer such discussions of hybrid modeling and exciting advances towards declarative modeling to later contributions.

\begin{acknowledgement}
The authors would like to acknowledge our collaborators on these works, Drs. Alexander Doroshenko, Karsten Hempel, Weicheng ``Winchell'' Qian, and Ellen Rafferty, as well as the Canadian Immunization Research Network (CIRN) and Mathematics for Public Health (MfPH).  Co-author Osgood wishes to express his appreciation of support from NSERC via the Discovery Grants program (RGPIN 2017-04647) and from SYK \& XZO.
\end{acknowledgement}

\bibliographystyle{spmpsci}
\bibliography{Project}

\end{document}